\documentclass[lettersize,journal]{IEEEtran}
\usepackage{amsmath,amsfonts}
\usepackage{algorithmic}
\usepackage{algorithm}
\usepackage{array}
\usepackage{subfigure}
\usepackage{textcomp}
\usepackage{stfloats}
\usepackage{url}
\usepackage{verbatim}
\usepackage{graphicx}
\usepackage{cite}
\usepackage{color} 
  \usepackage{xcolor}
  \usepackage{graphicx}
\usepackage{tikz}
\begin{document}

\title{A Novel Numerical Method for Relaxing the Minimal Configurations  of TOA-Based Joint Sensors and Sources Localization}

 \author{Faxian Cao, Yongqiang Cheng, Adil Mehmood Khan, Zhijing Yang and Yingxiu Chang
\thanks{This work was supported in part by China Scholarship Council ({\em Corresponding author: Yongqiang Cheng}).}
\thanks{F. Cao, A. M. Khan and Yingxiu Chang are with School of Computer Science, University of Hull, Hull HU6 7RX, U.K. (e-mail: faxian.cao-2022@hull.ac.uk; a.m.khan@hull.ac.uk; y.chang-2020@hull.ac.uk).}

\thanks{Y. Cheng is with Faculty of Technology, University of Sunderland, Sunderland SR6 0DD, U.K. (e-mail: yongqiang.cheng@sunderland.ac.uk).}

\thanks{Z. Yang is with School of Information Engineering, Guangdong University of Technology, Guangzhou 510006, China (e-mail:yzhj@gdut.edu.cn).}

}

\markboth{IEEE TRANSACTIONS ON...}%
{Shell \MakeLowercase{\textit{et al.}}: A Sample Article Using IEEEtran.cls for IEEE Journals}


\maketitle

\begin{abstract}
This work introduces a novel numerical method that relaxes the minimal configuration requirements for joint sensors and sources localization (JSSL) in 3D space using time of arrival (TOA) measurements. Traditionally, the principle requires that the number of valid equations (TOA measurements) must be equal to or greater than the number of unknown variables (sensor and source locations). State-of-the-art literatures suggest that the minimum numbers of sensors and sources needed for localization are four/five/six and six/five/four, respectively. However, these stringent configurations limit the application of JSSL in scenarios with an insufficient number of sensors and sources.
To overcome this limitation, we propose a numerical method that reduces the required number of sensors and sources, enabling more flexible JSSL configurations. First, we formulate the JSSL task as a series of triangles and apply the law of cosines to determine four unknown distances associated with one pair of sensors and three pairs of sources. Next, by utilizing triangle inequalities, we establish the lower and upper boundaries for these unknowns based on the known TOA measurements. The numerical method then searches within these boundaries to find the global optimal solutions, demonstrating that JSSL in 3D space is achievable with only four sensors and four sources—significantly relaxing the minimal configuration requirements.
Theoretical proofs and simulation results confirm the feasibility and effectiveness of the proposed method.
\end{abstract}


\section{Introduction}\label{sec:intr}
\IEEEPARstart{L}{ocalizing} both sources and receivers of sound signals is a central problem in signal processing due to its broad range of applications and its significance to many other critical challenges~\cite{sota-20}, such as noise reduction~\cite{Consensus-1} and signal separation~\cite{Consensus-2}. When sound sources emit audio signals and sensors capture these signals, time of arrival (TOA) or range measurements between sensors and sources can be obtained using cross-correlation techniques~\cite{gcc, phdrevis-1, phdrevis-2} or the mutual information function~\cite{phdrevis-3, phdrevis-4}, provided that the signals from both sensors and sources are synchronized (i.e., both share a common clock)~\cite{myself-1, myself-2}. If only the signals from sensors are available, the time difference of arrival (TDOA) can still be determined using cross-correlations~\cite{gcc} or the mutual information functions~\cite{phdrevis-3}. TDOA measures the range difference between a pair of sensors relative to the corresponding source, assuming that all sensors are synchronized~\cite{myself-1,myself-2}.

Localization methods using TOA or TDOA measurements in state-of-the-art approaches can be broadly categorized into two groups: closed-form (or near closed-form) solutions~\cite{sota-20, 31} and iterative methods~\cite{phdrevis-5, phdrevis-6}. For closed-form solutions,
given TDOA measurements and the positions of sensors, sources can be localized using techniques such as spherical interpolation,~\cite{phdrevis-7, phdrevis-8}, hyperbolic intersection method~\cite{phdrevis-9, phdrevis-10}, or linear intersection methods~\cite{phdrevis-11}.  Some applications~\cite{5} assume that all sensors and sources are co-located—-meaning a sensor is positioned very close to its corresponding source—-allowing for localization via multidimensional scaling with the distances between sensors/sources~\cite{5,sota-3}.  These types of localization problems have been extensively reviewed~\cite{phdrevis-14}, and their mathematical properties are well-documented~\cite{phdrevis-12, phdrevis-13}. In one approach, \textit{Crocco et al.}~\cite{4} present a closed-form solution for joint sensors and sources localization (JSSL) based on the low-rank property (LRP)~\cite{9} and TOA measurements, assuming a pair of co-located sensor and source.  Furthermore, \textit{Le et al.}~\cite{sota-20} apply LRP~\cite{9} and a linear method for solving polynomial equations~\cite{phdrevis-15, phdrevis-16} to derive closed-form and near closed-form solutions for JSSL with TOA measurements, specifically when the number of sensors and sources is greater than or equal to seven and four, respectively. With TDOA measurements, \textit{Le et al.}~\cite{mfaoa1} first recover TOA measurements from TDOA data, then present closed-form and near closed-form solutions~\cite{31} for JSSL when the number of sensors and sources is at least seven and five, respectively. However, their solutions do not hold for certain specific configurations~\cite{31}: (1) seven sensors and five sources, (2) seven sensors and six sources, and (3) eight sensors and five sources. \textit{Le et al.}~\cite{30} also propose an approximate and accurate algebraic solution for self-localization with TOA information by converting the problem into finding an upper triangular linear transformation matrix. Additionally, \textit{Pollefeys et al.}~\cite{sota-15} introduce a method for JSSL with TDOA information based on rank-5 factorization~\cite{sota-14}, effective when there are ten sensors and five sources. Beyond closed-form solutions~\cite{sota-20, 31}, some research has adopted iterative methods, such as gradient descent or auxiliary functions for nonlinear least-squares optimization. These iterative solutions are applicable for JSSL using TOA~\cite{phdrevis-5, simu1, sota-27} or TDOA measurements~\cite{phdrevis-6, sota-11, sota-1, sota-24, sota-29}.

More importantly,
by investigating  the rigid bipartite graphs (i.e., TOA measurements for JSSL), \textit{Bolker et al.}~\cite{Consensus-3} demonstrated that the minimal configurations for JSSL should consist of at least three sensors and three sources when both are in 2D space, provided that no three positions among the sensors and sources lie on a straight line.  (see Theorems 11 and 14 in \cite{Consensus-3}). Additionally, for scenarios where both sensors and sources are in 3D space, Theorems 11 and 12 in \textit{Bolker}'s work~\cite{Consensus-3} establish that the minimal configurations for localizing sensors and sources are six/five/four sensors and four/five/six sources, respectively, with the condition that no three and four positions among the sensors and sources lie on a line or plane, respectively. Due to the rigid mathematical property pertaining to the minimal configurations in \cite{Consensus-3}, those minimal configurations/cases have become the consensus for JSSL. The consensus can be derived by the principle that the number of valid equations (TOA measurements) should be at least equal to the number of unknown variables (locations of sensors and sources). 

To achieve the minimal configurations for JSSL without considering any constraints on the geometry of sensors and sources, many methods have been proposed in state-of-the-arts. For example, by studying the Grobner basis method~\cite{Consensus-5} and using the Macaulay2 software~\cite{Consensus-6} to solve polynomial equations~\cite{sota-20},  the closed-form solutions have been proposed by many researchers in Lund University~\cite{sota-17, sota-18, Consensus-4}. Specifically, when both sensors and sources are located in 3D, the closed-form solutions have been proposed by \textit{Kuang et al.}~\cite{sota-17}. When the span of sensors/sources locations has a higher dimension than the corresponding span of  sources/sensors locations, the closed-form solutions have been investigated by \textit{Burgess et al.}~\cite{sota-18}. \textit{Zhayida et al.}~\cite{Consensus-4} examined scenarios where two distances are provided between any one pair of sensors and any one pair of sources for the cases of four/five sensors and five/four sources. 



Despite the developments above for localizing both sensors and sources, a pressing and significant question has lingered: Can we relax the minimal configurations of state-of-the-arts for JSSL? Specifically, consider the challenging scenario where the number of sensors and sources is less than the minimal configurations stated in state-of-the-art. In such cases, there are neither iterative nor closed-form methods available for localization, which limits the applications of JSSL.   
However, if we can identify an alternative approach that bypasses the JSSL principle related to the number of valid equations and unknown locations, it may be possible to further reduce the number of sensors and sources needed for JSSL. Answering this question has profound implications for this field, as not only it has been unexplored, it can also make the configurations of JSSL more flexible since we can use fewer number of required sensors and sources for JSSL.  Thus, relaxing the minimal configurations for JSSL would be a landmark contribution to the development of JSSL techniques.

Accordingly, in this paper, we present a novel numerical method aimed at relaxing the minimal configurations required by state-of-the-art TOA measurements-based JSSL methods over the past decades, thereby facilitating the localization of both sensors and sources and making JSSL configurations more flexible.
We begin by formulating the JSSL problem as several triangles, using the laws of cosine. This allows us to transform the problem into the estimation of four unknown pairs of distances related to one pair of sensors and three pairs of sources. Consequently, the locations of all sensors and sources can be represented by only four unknown variables. Then, by applying triangle inequalities, we determine the lower and upper boundaries for these four unknown pairs of distance measurements based on known TOA measurements between sensors and sources. With these boundaries established, we search for the optimal solutions for the four unknowns by verifying candidates within the corresponding boundaries, using a small step size. Our proposed numerical method demonstrates that the minimal configurations for JSSL can be achieved with only four sensors and four sources, thus relaxing the minimal configurations required by state-of-the-art methods over the past decades. This leads to the conclusion that four sensors and four sources are sufficient for JSSL. Therefore, this study represents a significant advancement in the field of JSSL.   

\section{Problem Formulation}\label{sec:pro}
Consider a setup in 3D spaces with $M$ sensors $\begin{bmatrix}
    \mathbf{r}_1, \cdots,  \mathbf{r}_M
\end{bmatrix}_{3 \times M}$ and $N$ sources $\begin{bmatrix}
    \mathbf{s}_1, \cdots,  \mathbf{s}_N
\end{bmatrix}_{3 \times N}$. By defining the speed of sound as $c$, the TOA measurement between $i^{th}$ sensor and $j^{th}$ sources is given by~\cite{sota-20}
\begin{equation}
    \label{pfeq1}
    \mathbf{t}_{i,j}=\frac{\|\mathbf{r}_i-\mathbf{s}_j\|}{c},
\end{equation}
where $i=\begin{matrix}
    1, \cdots, M
\end{matrix}$, $j=\begin{matrix}
    1, \cdots, N
\end{matrix}$ and $\|\bullet\|$ denotes the $l_2$ norm. Due to the invariance of translation, rotation and reflection for the geometry of sensors and sources~\cite{sota-20}, we can assume $\mathbf{r}_1=\begin{bmatrix}
    0, & 0,& 0
\end{bmatrix}^T$, $\mathbf{r}_2=\begin{bmatrix}
    0, & 0,& \mathbf{r}_{3,2}
\end{bmatrix}^T$ and $\mathbf{s}_1=\begin{bmatrix}
    0, & \mathbf{s}_{2,1},& \mathbf{s}_{3,1}
\end{bmatrix}^T$, where $\mathbf{r}_{3,2}> 0$, $\mathbf{s}_{2,1}> 0$ and $[\ \bullet \ ]^T$ is the vector/matrix transpose. 


Upon inspection of Eq. (\ref{pfeq1}), when localizing all sensors and sources, it is obvious that the number of equations $MN$ (number of TOA measurements)  should be greater than or equal to the number of unknown variables $3(M+N)$. Considering the invariance of translation, rotation and reflection for the geometry of sensors and sources, sate-of-the-art methods over the past decades have established that the following inequality must hold for JSSL~\cite{Consensus-3}
\begin{equation}
    \label{pfeq3}
    MN\geq 3(M+N)-\frac{d(d+1)}{2},
\end{equation}
where $d=3$ denotes 3D space, thus $\frac{d(d+1)}{2}=6$ is the invariance of translation, rotation and reflection for the geometry of sensors and sources (see text below Eq. (\ref{pfeq1})). By performing some simple mathematical derivations, we can rewrite Eq. (\ref{pfeq3}) as
\begin{equation}
    \label{pfeq4}
    (M-3)(N-3)\geq 3.
\end{equation}

Upon inspection of Eq. (\ref{pfeq4}), it is obvious that the minimal configurations for JSSL should be: 1) $M=4$ and $N=6$, 2) $M=5$ and $N=5$ and 3) $M=6$ and $N=4$. This implies that the total number of sensors and sources should be at least $10$; otherwise, it is impossible to localize both sensors and sources. Over the past decades, the state-of-the-art methods have tried to approach these theoretical minimal configurations by presenting either closed-form or iterative solutions for JSSL. However, this limits the applications for TOA-based JSSL, especially when there are not sufficient number of devices/sensors and targets/sources for JSSL. 

Therefore, the objective of this paper is to relax the these theoretical minimal configurations established by the state-of-the-art methods. By doing so, we aim to make JSSL configurations more flexible, facilitating the tasks of JSSL by reducing the number of sensors and sources required. This represents a groundbreaking advancement in the the field of JSS, potentially revolutionizing JSSL techniques and expanding their applicability in more challenging environments.

\section{Preliminaries}\label{sec:pre}
This section shows the LRP~\cite{9} that was used for localizing both sensors and sources.

By taking the square of TOA formula of Eq. (\ref{pfeq1}) in both sides, we can have~\cite{sota-20}
\begin{equation}
    \label{pfeq5}
    \mathbf{r}_i^T\mathbf{r}_i - 2\mathbf{r}_i^T\mathbf{s}_j + \mathbf{s}_j^T\mathbf{s}_j =(c\mathbf{t}_{i,j})^2=\mathbf{d}_{i,j}^2,
\end{equation}
where $i=\begin{matrix}
    1, \cdots, M
\end{matrix}$ and $j=\begin{matrix}
    1, \cdots, N
\end{matrix}$, $\mathbf{d}_{i,j}$ is the range measurement between $i^{th}$ sensor and $j^{th}$ source.
Once we sequentially  subtract the equation for $j=1$ and the equation for $i=1$ and then add the equation for $i=j=1$ from Eq. (\ref{pfeq5}), we can have below equation~\cite{sota-20}
\begin{equation}
\label{pfeq6}
    (\mathbf{r}_i-\mathbf{r}_1)^T (\mathbf{s}_j-\mathbf{s}_1)=-\frac{1}{2}(\mathbf{d}_{i,j}^2-\mathbf{d}_{i,1}^2-\mathbf{d}_{1,j}^2+\mathbf{d}_{1,1}^2).
\end{equation}
where $i=\begin{matrix}
    2, \cdots, M
\end{matrix}$ and $j=\begin{matrix}
    2, \cdots, N
\end{matrix}$. Upon inspection of Eq. (\ref{pfeq6}), by denoting $\mathbf{R}_{:,i-1}=\mathbf{r}_i-\mathbf{r}_1$, $\mathbf{S}_{:,j-1}=\mathbf{s}_j-\mathbf{s}_1$ and $\mathbf{D}_{i-1,j-1}=\mathbf{d}_{i,j}^2-\mathbf{d}_{i,1}^2-\mathbf{d}_{1,j}^2+\mathbf{d}_{1,1}^2$ where $i=\begin{matrix}
    2, \cdots, M
\end{matrix}$ and $j=\begin{matrix}
    2, \cdots, N
\end{matrix}$, Eq. (\ref{pfeq6}) can be re-written as matrix form~\cite{sota-20}
\begin{equation}
    \label{pfeq7}
    -2\mathbf{R}^T\mathbf{S}=\mathbf{D},
\end{equation}
where $\mathbf{R} \in \mathbb{R}^{3 \times (M-1)}$ and $\mathbf{S} \in \mathbb{R}^{3 \times (N-1)}$.

From Eq. (\ref{pfeq7}), we observe that the left side involves the unknown positions of both the sensors and the sources, whereas the right side consists of the known TOA measurements. Moreover, since both sensors and sources are situated in a 3D space, the LRP~\cite{9} for these elements can be expressed as
\begin{equation}
    \label{pfeq8}
    rank(\mathbf{R}^T\mathbf{S})=rank(\mathbf{D})\leq 3.
\end{equation}


By applying the singular value decomposition~\cite{phdrevis-17} to matrix $\mathbf{D}$ in Eq. (\ref{pfeq8}), we can have
\begin{equation}
    \label{preq1}
    \mathbf{D}=\mathbf{\hat{U}} \mathbf{\hat{A}} \mathbf{\hat{V}}^T,
\end{equation}
where $\mathbf{\hat{U}}\in \mathbb{R}^{(M-1) \times (M-1)}$ and $\mathbf{\hat{V}}^T\in \mathbb{R}^{(N-1) \times (N-1)}$ are the left and right singular matrices, respectively, and $\mathbf{\hat{A}}\in \mathbb{R}^{(M-1) \times (N-1)}$ is the corresponding singular values. Then from the LRP $rank(\mathbf{D})\leq 3$ in Eq. (\ref{pfeq8}),
by defining $\mathbf{U}=\mathbf{\hat{U}}_{:,1:3}\in \mathbb{R}^{(M-1) \times 3}$, $\mathbf{V}=\mathbf{\hat{A}}_{1:3,1:3} \mathbf{\hat{V}}_{:,1:3}^T\in \mathbb{R}^{3 \times (N-1)}$, Eq. (\ref{preq1}) can be re-written as~\cite{sota-20}
\begin{equation}
    \label{preq2}
    \mathbf{D}=\mathbf{U}\mathbf{V}.
\end{equation}

Finally, with  Eqs. (\ref{pfeq7}) and (\ref{preq2}), by defining an unknown matrix $\mathbf{C}\in \mathbb{R}^{3 \times 3}$, we can have~\cite{sota-20}
\begin{equation}
    \label{preq3}
    \begin{cases}
        \mathbf{R}^T=\mathbf{U}\mathbf{C}^{-1} \\
        -2\mathbf{S}=\mathbf{C}\mathbf{V}
    \end{cases}.
\end{equation}

Upon inspection of Eqs. (\ref{pfeq6}), (\ref{pfeq7}) and (\ref{preq3}), since $\mathbf{r}_1=\begin{bmatrix}
       0, & 0, & 0
\end{bmatrix}^T$ and $\mathbf{s}_1=\begin{bmatrix}
       0, & \mathbf{s}_{2,1}, & \mathbf{s}_{3,1}
\end{bmatrix}^T$,
it is obvious that once we can obtain the corresponding eleven variables (two variables for $\mathbf{s}_{2,1}$ and  $\mathbf{s}_{3,1}$ and 
nine variables in unknown matrix $\mathbf{C}$), the solutions for JSSL can be obtained.


With $M$ sensors and $N$ sources in 3D space, if any three/four sensors and sources are not co-linear/coplanar, based on the rigid bipartite graphs principle in 3D space~\cite{rigidbi}, the number of edges (representing distances between sensors and sources) must be at least $3(M+N)-6$. This ensures that the shapes formed by $M$ sensors and $N$ sources remain unchanged, indicating potential unique solutions for their locations. Hence, if there are $MN$ known TOA measurements between sensors and sources, a prerequisite for obtaining solutions is $MN \geq 3(M+N)-6$. This requirement aligns with the notion from Eq. (\ref{pfeq3}) that the number of valid equations ($MN$ known distances) should exceed or equal the number of unknowns (with $3(M+N)-6$ locations for $M$ sensors and $N$ sources). This consensus highlights the necessity of having a minimum of four/five/six sensors and six/five/four sources for JSSL. However, in scenario where there are insufficient sensors/devices or sources/targets for JSSL, this requirement restricts the flexibility of configurations for JSSL, making the task of JSSL failure. Therefore, if we can relax the minimal configurations suggested by state-of-the-art in the past decades, allowing fewer number of sensors and sources for JSSL, this increased flexibility in sensor and source configurations has the potential to revolutionize the techniques of JSSL.


\section{Proposed Numerical Method Based On Triangles}\label{sec:prosed}

In this section, based on the properties of triangles, such as laws of cosine and triangle inequality, we propose a new numerical method to reduce the number of sensors and sources for JSSL, relaxing the minimal configurations for JSSL, enabling more flexible configurations for JSSL, so that we can facilitate the task of JSSL in more challenging environments. 
Next, in Subsection \ref{sec:prosed-1}, by constructing several triangles, we transform the localization problems of both sensors and sources to obtain the solutions of four unknown distances pertaining to one pair of sensors and three pairs of sources. Then the effects of those four unknowns on locations of sensors and sources are illustrated in  Subsection \ref{sec:prosed-2}. In Subsections \ref{sec:prosed-3} and \ref{sec:prosed-4}, to obtain the numerical solutions of four unknowns for localizing both sensors and sources, we first present the lower and upper boundaries for the corresponding four unknowns in  Subsection \ref{sec:prosed-3} based on triangle inequality, so that the numerical solutions of four unknowns can be obtained by searching the candidates within the corresponding boundaries (see Subsection \ref{sec:prosed-4}).  Finally, we extend the proposed numerical solutions to several scenarios of JSSL in Subsection \ref{sec:prosed-5}, showing the flexibility of proposed numerical method on the task of JSSL.

\begin{figure*}[!ht]
\centering
\subfigure[Triangle for $r_1$, $r_2$ and $s_j$.]{\includegraphics[trim=0.2cm 0.4cm 1.3cm 0.5cm, clip=false, width=0.22\linewidth]{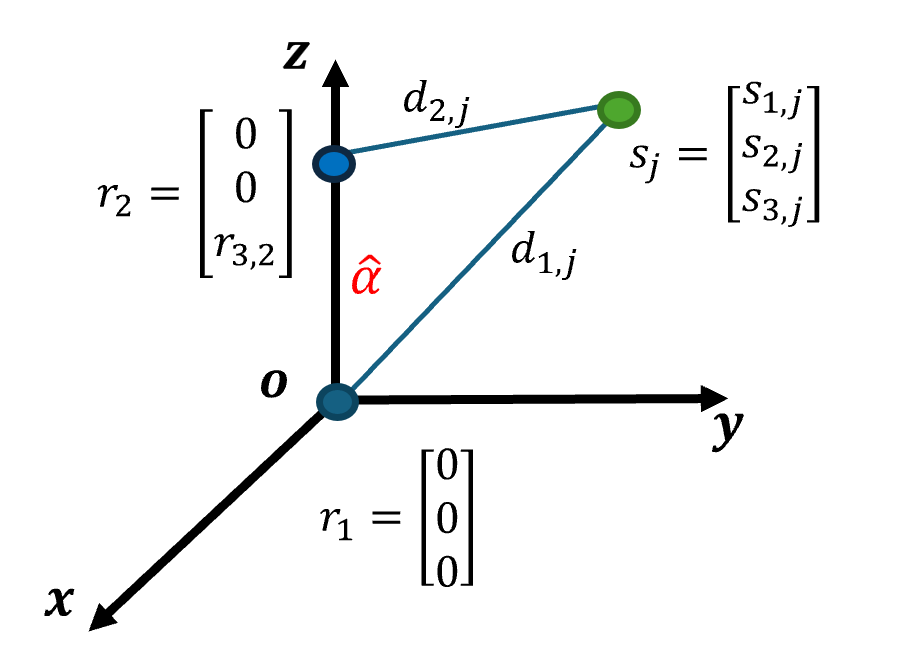}}
\subfigure[Triangle for $r_1$, $s_1$ and $s_2$.]{\includegraphics[trim=0.2cm 0.4cm 1.3cm 0.5cm, clip=false,width=0.22\linewidth]{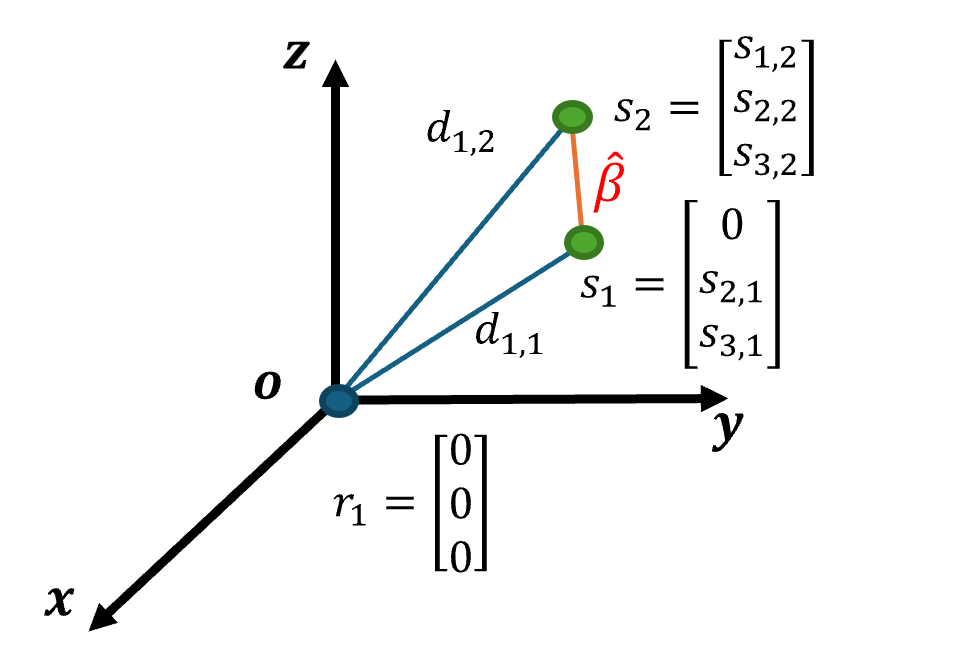}}
\subfigure[Triangle for $r_1$, $s_1$ and $s_3$.]{\includegraphics[trim=0.2cm 0.4cm 1.3cm 0.5cm, clip=false,width=0.22\linewidth]{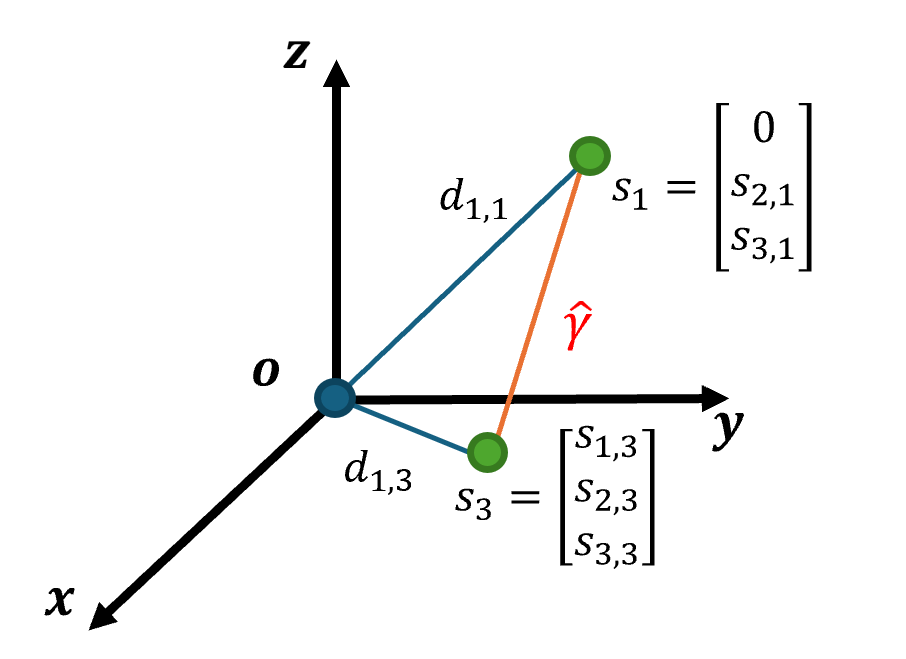}}
\subfigure[Triangle for $r_1$, $s_1$ and $s_4$.]{\includegraphics[trim=0.2cm 0.4cm 1.3cm 0.5cm, clip=false,width=0.22\linewidth]{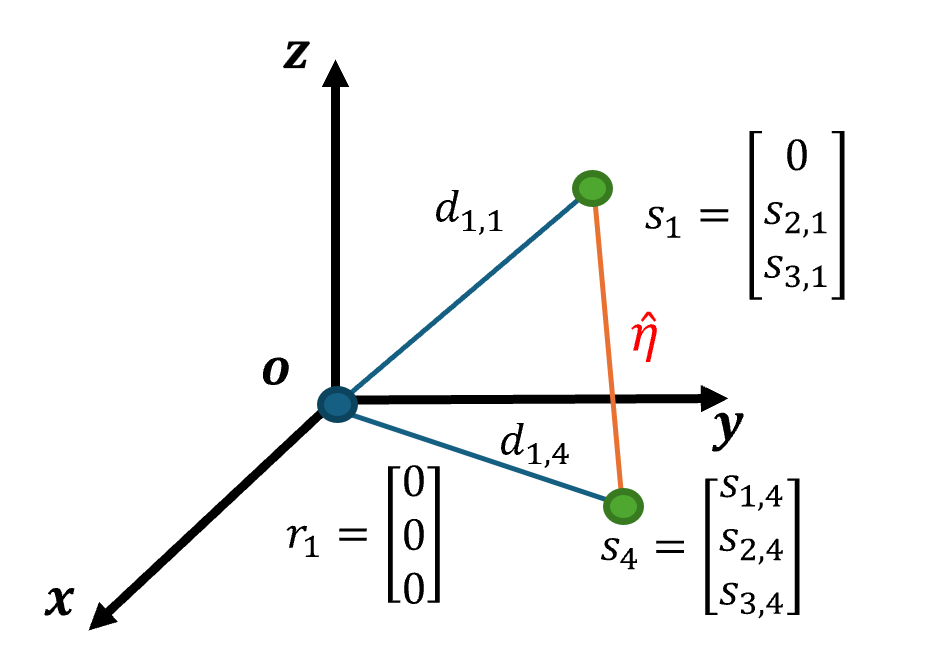}}

\caption{Triangles for transforming JSSL to four unknown pairs of distance measurements.}
\label{over-fig1}
\end{figure*}
\subsection{Laws of cosine for transformations of JSSL}\label{sec:prosed-1}

Upon inspection of Eq. (\ref{preq3}), it is obvious that
\begin{equation}
    \label{Con-1}
    \begin{cases}
        (\mathbf{r}_i-\mathbf{r}_1)^T=\mathbf{U}_{i-1,:}\mathbf{C}^{-1} \\
        -2(\mathbf{s}_j-\mathbf{s}_1)= \mathbf{C} \mathbf{V}_{:,j-1}
    \end{cases},
\end{equation}
where $i=\begin{matrix}
    2, & \cdots, & M
\end{matrix}$ and $j=\begin{matrix}
    2, & \cdots, & N
\end{matrix}$. Since $\mathbf{r}_1=\begin{bmatrix}
     0, & 0, & 0
 \end{bmatrix}^T$, $\mathbf{r}_2=\begin{bmatrix}
     0, & 0, & \mathbf{r}_{3,2}
 \end{bmatrix}^T$ and $\mathbf{s}_1=\begin{bmatrix}
    0, & \mathbf{s}_{2,1}, & \mathbf{s}_{3,1}
 \end{bmatrix}^T$, respectively, where $\mathbf{r}_{3,2}>0$ and $\mathbf{s}_{2,1}>0$, it is obvious that we can localize all sensors and sources once the solutions for $\mathbf{s}_{2,1}$ and $\mathbf{s}_{3,1}$ as well as the nine unknown variables in matrix $\mathbf{C}$ are estimated. Next, we will formulate the solutions of those eleven variables as four unknown pairs of distance measurements pertaining to a pair of sensors and three pairs of sources. By defining $\hat{\alpha}$ as the distance between the $1^{st}$ and $2^{nd}$ sensors, $\hat{\beta}$ as the distance between the $1^{st}$ and $2^{nd}$ sources, $\hat{\gamma}$ as the distance between the $1^{st}$ and $3^{rd}$ sources, $\hat{\eta}$ as the distance between the $1^{st}$ and $4^{th}$ sources, then from Fig. \ref{over-fig1}(a), we can have the corresponding equation by using laws of cosine
 \begin{equation}
   \label{Con-2}
\hat{\alpha}^2=\mathbf{d}_{1,j}^2+\mathbf{d}_{2,j}^2-2(\mathbf{r}_1-\mathbf{s}_j)^T(\mathbf{r}_2-\mathbf{s}_j),
\end{equation}
where $(r_1-s_j)^T(r_2-s_j)=\|r_1-s_j\|\|r_2-s_j\| 
 \ cos(\theta_{r_1-s_j, r_2-s_j})$ and $\theta_{r_1-s_j, r_2-s_j}$ is the angle between vectors $r_1-s_j$ and $r_2-s_j$.

Upon inspection of Eq. (\ref{Con-2}),
with  $\mathbf{r}_1=\begin{bmatrix}
     0, & 0, & 0
 \end{bmatrix}^T$, $\mathbf{r}_2=\begin{bmatrix}
     0, & 0, & \mathbf{r}_{3,2}
 \end{bmatrix}^T$ and $(\mathbf{r}_1-\mathbf{s}_j)^T(\mathbf{r}_1-\mathbf{s}_j)=\mathbf{d}_{1,j}^2$, where $\mathbf{r}_{3,2}=\hat{\alpha}$ and $j$ ranges from $1$ to $N$, it implies
 \begin{equation}
       \label{Con-3}    \mathbf{s}_{3,j}=\frac{\hat{\alpha}^2+\mathbf{d}_{1,j}^2-\mathbf{d}_{2,j}^2}{2\hat{\alpha}},
 \end{equation}
where $j=\begin{matrix}
    1, & \cdots, & N
\end{matrix}$.

Upon inspection of Eq. (\ref{Con-3}) and $(\mathbf{r}_1-\mathbf{s}_1)^T(\mathbf{r}_1-\mathbf{s}_1)=\mathbf{d}_{1,1}^2$, since $\mathbf{s}_{1,1}=0$ and $\mathbf{s}_{2,1}>0$, we can have  
\begin{equation}
      \label{Con-4}
      \mathbf{s}_{2,1}=\sqrt{\mathbf{d}_{1,1}^2-\mathbf{s}_{3,1}^2}.
\end{equation}

With Eqs. (\ref{Con-3}) and (\ref{Con-4}), it is observed that two unknown variables $\mathbf{s}_{2,1}$ and $\mathbf{s}_{3,1}$ can be represented by the unknown distance $\hat{\alpha}$ between $1^{st}$ and $2^{nd}$ sensors, thus, the following task is to obtain the solutions of nine variables in matrix $\mathbf{C}$. From Eq. (\ref{Con-1}), we can rewrite $ -2(\mathbf{s}_j-\mathbf{s}_1)= \mathbf{C}\mathbf{V}_{:,j}$ as
\begin{equation}
    \label{Con-5}
    \mathbf{C}=-2\begin{bmatrix}
        \mathbf{s}_2-\mathbf{s}_1 & \mathbf{s}_3-\mathbf{s}_1 & \mathbf{s}_4-\mathbf{s}_1
    \end{bmatrix} 
        \mathbf{V}_{:,1:3}^{-1},
\end{equation}
thus the three unknown variables in the third row of matrix $C$ are
\begin{equation}
    \label{Con-6}
    \mathbf{C}_{3,:}=-2\begin{bmatrix}
        \mathbf{s}_{3,2}-\mathbf{s}_{3,1} & \mathbf{s}_{3,3}-\mathbf{s}_{3,1} & \mathbf{s}_{3,4}-\mathbf{s}_{3,1}
    \end{bmatrix} 
        \mathbf{V}_{:,1:3}^{-1}.
\end{equation}

By applying Eq. (\ref{Con-3}) to Eq. (\ref{Con-6}), we have
\begin{align}
     \label{Con-7}
     \mathbf{C}_{3,:}&=-\frac{\begin{bmatrix}
        \mathbf{d}_{1,2}^2-\mathbf{d}_{2,2}^2-\mathbf{d}_{1,1}^2+\mathbf{d}_{2,1}^2 \\  \mathbf{d}_{1,3}^2-\mathbf{d}_{2,3}^2-\mathbf{d}_{1,1}^2+\mathbf{d}_{2,1}^2 \\  \mathbf{d}_{1,4}^2-\mathbf{d}_{2,4}^2-\mathbf{d}_{1,1}^2+\mathbf{d}_{2,1}^2
    \end{bmatrix}^T \mathbf{V}_{:,1:3}^{-1}}{\hat{\alpha}} \nonumber\\
    &= \frac{\mathbf{D}_{1,1:3}\mathbf{V}_{:,1:3}^{-1}}{\hat{\alpha}} 
    =\frac{\mathbf{U}_{1,:}}{\hat{\alpha}}.    
\end{align}

Upon inspection of Eq. (\ref{Con-7}),  since matrix $\mathbf{U}$ is known, it is obvious that the three variables in the third row of matrix $\mathbf{C}$ are the functions of unknown $\hat{\alpha}$, thus, we can observe that the five unknown variables, $\mathbf{s}_{2,1}$, $\mathbf{s}_{3,1}$ and $\mathbf{C}_{3,:}$ are the functions of $\hat{\alpha}$ only by inspecting Eqs. (\ref{Con-3}), (\ref{Con-4}) and (\ref{Con-7}). Next, our remaining task is to have the solutions of remaining six unknown variables in matrix $\mathbf{C}$.

From Figs. \ref{over-fig1}(b), (c) and (d), by applying the laws of cosine to the corresponding three triangles, we can have
\begin{equation}
    \label{Con-8}
    \begin{cases}
        \hat{\beta}^2=\mathbf{d}_{1,1}^2+\mathbf{d}_{1,2}^2-2\mathbf{s}_1^T\mathbf{s}_2 \\
         \hat{\gamma}^2=\mathbf{d}_{1,1}^2+\mathbf{d}_{1,3}^2-2\mathbf{s}_1^T\mathbf{s}_3 \\
         \hat{\eta}^2=\mathbf{d}_{1,1}^2+\mathbf{d}_{1,4}^2-2\mathbf{s}_1^T\mathbf{s}_4
    \end{cases}.
\end{equation}

Since $\mathbf{s}_j^T\mathbf{s}_j=\mathbf{d}_{1,j}^2$, we can rewrite Eq. (\ref{Con-8}) as
\begin{equation}
    \label{Con-9}
    \begin{cases}
        \hat{\beta}^2=\mathbf{d}_{1,2}^2-\mathbf{d}_{1,1}^2-2\mathbf{s}_1^T(\mathbf{s}_2-\mathbf{s}_1) \\
         \hat{\gamma}^2=\mathbf{d}_{1,3}^2-\mathbf{d}_{1,1}^2-2\mathbf{s}_1^T(\mathbf{s}_3-\mathbf{s}_1) \\
         \hat{\eta}^2=\mathbf{d}_{1,4}^2-\mathbf{d}_{1,1}^2-2\mathbf{s}_1^T(\mathbf{s}_4-\mathbf{s}_1)
    \end{cases}.
\end{equation}

Upon inspection of Eq. (\ref{Con-9}), by applying both $-2(\mathbf{s}_j-\mathbf{s}_1)=\mathbf{C}\mathbf{V}_{:,j-1}$ and $\mathbf{s}_1=\begin{bmatrix}
    0, & \mathbf{s}_{2,1}, & \mathbf{s}_{3,1}
\end{bmatrix}^T$ to Eq. (\ref{Con-9}), it follows
\begin{equation}
    \label{Con-10}
        \begin{cases}
        \hat{\beta}^2=\mathbf{d}_{1,2}^2-\mathbf{d}_{1,1}^2+\mathbf{s}_{2,1}\mathbf{C}_{2,:}\mathbf{V}_{:,1}+\mathbf{s}_{3,1}\mathbf{C}_{3,:}\mathbf{V}_{:,1}  \\
         \hat{\gamma}^2=\mathbf{d}_{1,3}^2-\mathbf{d}_{1,1}^2+\mathbf{s}_{2,1}\mathbf{C}_{2,:}\mathbf{V}_{:,2}+\mathbf{s}_{3,1}\mathbf{C}_{3,:}\mathbf{V}_{:,2} \\
         \hat{\eta}^2=\mathbf{d}_{1,4}^2-\mathbf{d}_{1,1}^2+\mathbf{s}_{2,1}\mathbf{C}_{2,:}\mathbf{V}_{:,3}+\mathbf{s}_{3,1}\mathbf{C}_{3,:}\mathbf{V}_{:,3}
    \end{cases}.
\end{equation}

From Eq. (\ref{Con-10}), we can express the three variable in the second row of matrix $\mathbf{C}$ as
\begin{equation}
      \label{Con-11}
      \mathbf{C}_{2,:}=\frac{\begin{bmatrix}
          \hat{\beta}^2-\mathbf{d}_{1,2}^2+\mathbf{d}_{1,1}^2 \\ \hat{\gamma}^2-\mathbf{d}_{1,3}^2+\mathbf{d}_{1,1}^2 \\ 
          \hat{\eta}^2-\mathbf{d}_{1,4}^2+\mathbf{d}_{1,1}^2
      \end{bmatrix}^T\mathbf{V}_{:,1:3}^{-1}-\mathbf{s}_{3,1}\mathbf{C}_{3,:}}{\mathbf{s}_{2,1}}.
\end{equation}

Upon inspection of Eq. (\ref{Con-11}), since $\mathbf{s}_{2,1}$, $\mathbf{s}_{3,1}$ and $\mathbf{C}_{3,:}$ are the functions of $\hat{\alpha}$, it implies the three variables $\mathbf{C}_{2,:}$ are the functions of four unknown distances $\hat{\alpha}$, $\hat{\beta}$, $\hat{\gamma}$ and $\hat{\eta}$. Thus, from the solutions of $\mathbf{C}_{2,:}$ in Eq. (\ref{Con-11}), it is obvious that once $\mathbf{r}_1$, $\mathbf{r}_2$ and  $\mathbf{s}_1$ are co-linear, the solutions of $\mathbf{C}_{2,:}$ are  not stable if there are estimation errors for $\hat{\alpha}$, $\hat{\beta}$, $\hat{\gamma}$ and $\hat{\eta}$. However, as we mentioned earlier, given any three/four positions among sensors and sources, they should not be lying in a same line/plane. Therefore, this special case for the solution of $\mathbf{C}_{2,:}$ can be excluded.  Finally, once we can represent the remaining three variables $\mathbf{C}_{1,:}$ as the functions of four unknown distances $\hat{\alpha}$, $\hat{\beta}$, $\hat{\gamma}$ and $\hat{\eta}$, the locations of all sensors and sources can be represented by four unknown distance measurements above.

Considering $\mathbf{s}_j^T\mathbf{s}_j=\mathbf{d}_{1,j}^2$ (see Fig. \ref{over-fig1}(a)), we can have
\begin{equation}
    \label{Con-12}
    \begin{cases}
        \mathbf{d}_{1,2}^2=\mathbf{s}_{1,2}^2+\mathbf{s}_{2,2}^2+\mathbf{s}_{3,2}^2 \\
        \mathbf{d}_{1,3}^2=\mathbf{s}_{1,3}^2+\mathbf{s}_{2,3}^2+\mathbf{s}_{3,3}^2 \\
        \mathbf{d}_{1,4}^2=\mathbf{s}_{1,4}^2+\mathbf{s}_{2,4}^2+\mathbf{s}_{3,4}^2
    \end{cases}.
\end{equation}

To represent three variables in $\mathbf{C}_{1,:}$ as the functions of $\hat{\alpha}$, $\hat{\beta}$, $\hat{\gamma}$, $\hat{\eta}$, we can rewrite Eq. (\ref{Con-12}) as
\begin{small}
\begin{equation}
    \label{Con-13}
        \begin{cases}
      (\mathbf{s}_{1,2}-\mathbf{s}_{1,1}+\mathbf{s}_{1,1})^2+(\mathbf{s}_{2,2}-\mathbf{s}_{2,1}+\mathbf{s}_{2,1})^2 = \mathbf{d}_{1,2}^2-\mathbf{s}_{3,2}^2 \\
        (\mathbf{s}_{1,3}-\mathbf{s}_{1,1}+\mathbf{s}_{1,1})^2+(\mathbf{s}_{2,3}-\mathbf{s}_{2,1}+\mathbf{s}_{2,1})^2 = \mathbf{d}_{1,3}^2-\mathbf{s}_{3,3}^2 \\
            (\mathbf{s}_{1,4}-\mathbf{s}_{1,1}+\mathbf{s}_{1,1})^2+(\mathbf{s}_{2,4}-\mathbf{s}_{2,1}+\mathbf{s}_{2,1})^2 = \mathbf{d}_{1,4}^2-\mathbf{s}_{3,4}^2
    \end{cases},
\end{equation}
\end{small}
then by applying $-2(\mathbf{s}_j-\mathbf{s}_1)=\mathbf{C}\mathbf{V}_{j-1}$ and $\mathbf{s}_{1,1}=0$ to Eq. (\ref{Con-13}), where $j$ ranges from 2 to $N$, it follows
\begin{equation}
     \label{Con-14}
     \begin{cases}
        (\mathbf{C}_{1,:}\mathbf{V}_{:,1})^2=4(\mathbf{d}_{1,2}^2-\mathbf{s}_{3,2}^2)-(\mathbf{C}_{2,:}\mathbf{V}_{:,1}-2\mathbf{s}_{2,1})^2 \\
         (\mathbf{C}_{1,:}\mathbf{V}_{:,2})^2=4(\mathbf{d}_{1,3}^2-\mathbf{s}_{3,3}^2)-(\mathbf{C}_{2,:}\mathbf{V}_{:,2}-2\mathbf{s}_{2,1})^2   \\
         (\mathbf{C}_{1,:}\mathbf{V}_{:,3})^2=4(\mathbf{d}_{1,4}^2-\mathbf{s}_{3,4}^2)-(\mathbf{C}_{2,:}\mathbf{V}_{:,3}-2\mathbf{s}_{2,1})^2       
     \end{cases}.
\end{equation}

Upon inspection of Eq. (\ref{Con-14}), we can represent $\mathbf{C}_{1,:}$ as
\begin{equation}
    \label{Con-15}
   \mathbf{C}_{1,:} =\begin{bmatrix}
        \pm\sqrt{4(\mathbf{d}_{1,2}^2-\mathbf{s}_{3,2}^2)-(\mathbf{C}_{2,:}\mathbf{V}_{:,1}-2\mathbf{s}_{2,1})^2} \\ \pm\sqrt{4(\mathbf{d}_{1,3}^2-\mathbf{s}_{3,3}^2)-(\mathbf{C}_{2,:}\mathbf{V}_{:,2}-2\mathbf{s}_{2,1})^2  } \\
        \pm\sqrt{4(\mathbf{d}_{1,4}^2-\mathbf{s}_{3,4}^2)-(\mathbf{C}_{2,:}\mathbf{V}_{:,3}-2\mathbf{s}_{2,1})^2}
    \end{bmatrix}^T \mathbf{V}_{:,1:3}^{-1}.
\end{equation}

From Eq. (\ref{Con-15}), it is obvious that $\mathbf{C}_{1,:}$ can be represented by four unknown distance measurements $\hat{\alpha}$, $\hat{\beta}$, $\hat{\gamma}$ and $\hat{\eta}$ since $\mathbf{s}_{3,j}$, $\mathbf{s}_{2,1}$ are the functions of $\hat{\alpha}$ (see Eqs. (\ref{Con-3}) and (\ref{Con-4})) and the three variables of $\mathbf{C}_{2,:}$ are the functions of $\hat{\alpha}$, $\hat{\beta}$, $\hat{\gamma}$ and $\hat{\eta}$ (see Eq. (\ref{Con-11})). In addition, Eq. (\ref{Con-15}) shows the eight ambiguities for the solutions of $\mathbf{C}_{1,:}$, resulting in the additional difficulties for obtaining the solutions of $\mathbf{C}_{1,:}$. Fortunately, due to the invariance of reflection pertaining to the geometry of sensors and sources, eight ambiguities in Eq. (\ref{Con-15}) can be reduced to four ambiguities, i.e., $\{+, +, +\}$, $\{+, +, -\}$, $\{+, -, +\}$ and $\{+, -, -\}$. 

From Eqs. (\ref{Con-3}), (\ref{Con-4}) and (\ref{Con-7}), it is observed that the five variables, $\mathbf{s}_{2,1}$, $\mathbf{s}_{3,1}$ and $\mathbf{C}_{3,:}$, are the function of $\hat{\alpha}$. In addition, upon inspection of Eqs. (\ref{Con-11}) and (\ref{Con-15}), it is obvious that six variables in  $\mathbf{C}_{1,:}$ and $\mathbf{C}_{2,:}$ are the functions of $\hat{\alpha}$, $\hat{\beta}$, $\hat{\gamma}$ and $\hat{\eta}$. Therefore, the eleven variables above are the functions of  $\hat{\alpha}$, $\hat{\beta}$, $\hat{\gamma}$ and $\hat{\eta}$, it implies that the locations of all sensors and sources can be dominated by four unknown variables $\hat{\alpha}$, $\hat{\beta}$, $\hat{\gamma}$ and $\hat{\eta}$ only. Besides, it is obvious that there are four ambiguities in $\mathbf{C}_{1,:}$, however, those four ambiguities can be eliminated by comparing the distance errors between sensors and sources. In more details, once the solutions for four unknown variables $\hat{\alpha}$, $\hat{\beta}$, $\hat{\gamma}$ and $\hat{\eta}$ are estimated, then given four ambiguities in $\mathbf{C}_{1,:}$, we can obtain four different sets of locations for sensors and sources, resulting in four sets of distances between sensors and sources $\mathbf{d}_{i,j}^{(se)}=\|\mathbf{r}_i^{(se)}- \mathbf{s}_j^{(se)}\|$ ($se=\begin{matrix}
    1, & \cdots, & 4
\end{matrix}$, $i=\begin{matrix}
    1, & \cdots, & M
\end{matrix}$, $j=\begin{matrix}
    1, & \cdots, & N
\end{matrix}$), so that the ambiguities in $\mathbf{C}_{1,:}$ can be eliminated by comparing the errors between  ground truth of range measurements $\mathbf{d}_{i,j}$ and four sets of distance $\mathbf{d}_{i,j}^{(se)}$, i.e., $min\{\sqrt{\sum_{i=1}^{M}\sum_{j=1}^{N}(\mathbf{d}_{i,j}-\mathbf{d}_{i,j}^{(se)})^2}\}$, where $se=\begin{matrix}
    1, & \cdots, & 4 
\end{matrix}$. In the next subsection, we will display the effect of four unknowns $\hat{\alpha}$, $\hat{\beta}$, $\hat{\gamma}$ and $\hat{\eta}$ on the locations of sensors of sources.

\subsection{Effects of four unknowns on locations of sensors and sources}\label{sec:prosed-2}

\begin{figure*}[!t]
\centering
\subfigure[Effect of $\hat{\alpha}$ ($j\geq2$).]{\includegraphics[trim=0.2cm 0.4cm 1.3cm 0.5cm, clip=false, width=0.3\linewidth]{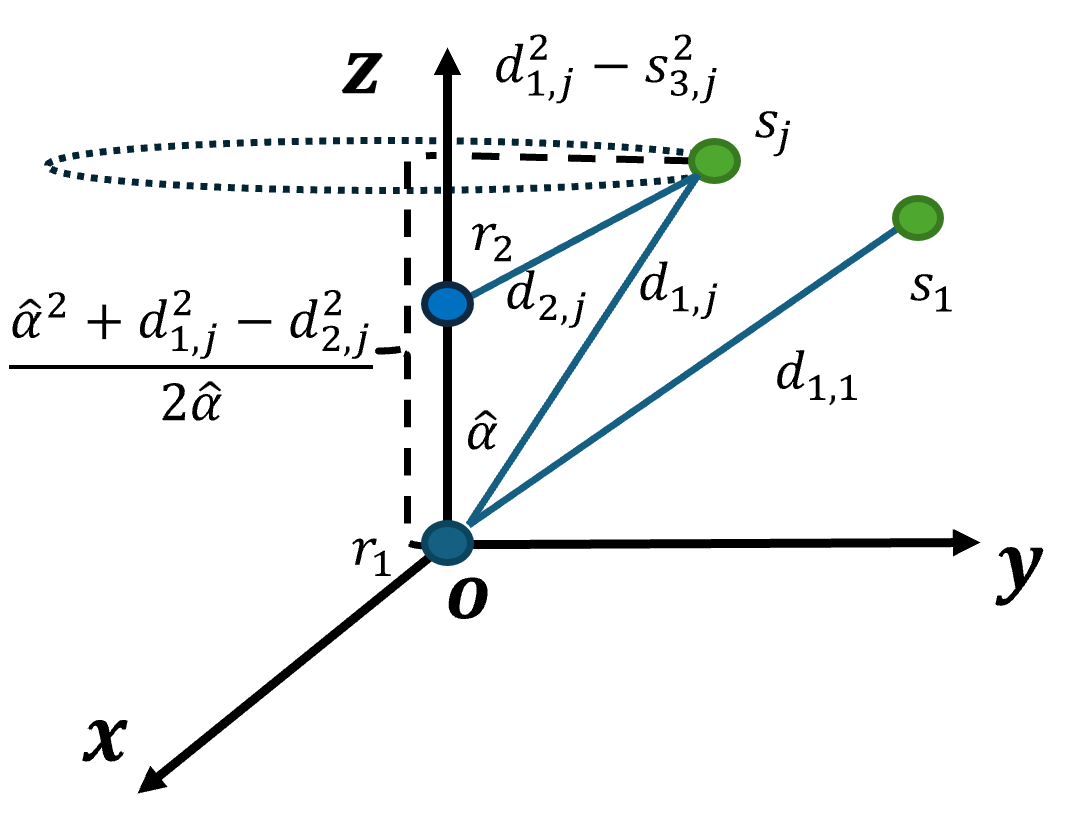}}
\subfigure[Effects of $\hat{\beta}$, $\hat{\gamma}$ and  $\hat{\eta}$.]{\includegraphics[trim=0.2cm 0.4cm 1.3cm 0.5cm, clip=false,width=0.3\linewidth]{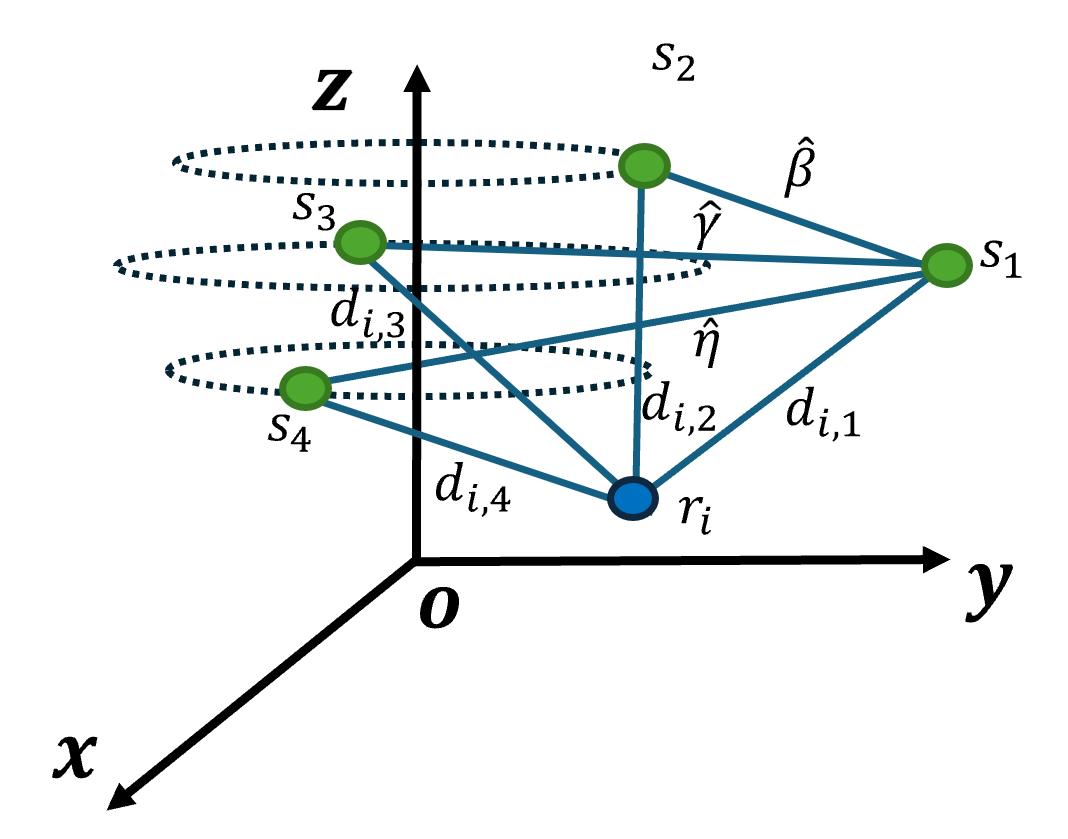}}
\caption{Effects of four unknowns on locations of sensors and sources.}
\label{Con-fig2}
\end{figure*}

Upon inspection of Eq. (\ref{Con-3}) and Fig. \ref{Con-fig2}(a), it is obvious that the third coordinate of all sources $\mathbf{s}_{3,j}$ can be dominated by the solution of unknown variable $\hat{\alpha}$, so that we can confirm that all possible solutions for the first and second coordinates of all sources  are lying in different circles, where the corresponding circle for each source can be dominated by $\hat{\alpha}$ since $\mathbf{s}_{1,j}^2+\mathbf{s}_{2,j}^2=\mathbf{d}_{1,j}^2-\mathbf{s}_{3,j}^2$. In addition, as the first and second coordinates of first source, $\mathbf{s}_{1,1}$ and $\mathbf{s}_{2,1}$, are zero and larger than zero, respectively, so that the location of first source $\mathbf{s}_{1}$ can be  dominated by $\hat{\alpha}$ only (see Eqs. (\ref{Con-3}) and (\ref{Con-4})). 

Besides, as confirmed by Eqs. (\ref{Con-7}), (\ref{Con-11}) and  (\ref{Con-15}), the nine variable in matrix $\mathbf{C}$ can be dominated by four unknown variables $\hat{\alpha}$, $\hat{\beta}$, $\hat{\gamma}$ and $\hat{\eta}$, resulting in the solutions for the locations of all sensors and sources. In more details, from Fig. \ref{Con-fig2}(b), it is obvious that once $\hat{\beta}$, $\hat{\gamma}$ and $\hat{\eta}$ are confirmed, we can localize three sources $\mathbf{s}_2$, $\mathbf{s}_3$ and $\mathbf{s}_4$ since the distance $\mathbf{d}_{i,j}$ between sensor $\mathbf{r}_i$ and source $\mathbf{s}_j$ is known. Thus, with the locations of $\mathbf{s}_{1}$, $\mathbf{s}_2$, $\mathbf{s}_3$ and $\mathbf{s}_4$, the solutions of eleven variables $\mathbf{s}_{2,1}$ $\mathbf{s}_{3,1}$, $\mathbf{C}$ can be obtained, resulting in the solutions for localizing all sensors and sources (see Eq. (\ref{Con-1})). Therefore, the estimation of four unknowns $\hat{\alpha}$, $\hat{\beta}$, $\hat{\gamma}$ and $\hat{\eta}$ becomes crucial for JSSL. Unfortunately, no prior information is available regarding the values of these four unknowns, making their estimation particularly challenging. Specifically, when there are only four sensors and four sources, the number of valid equations in $\mathbf{d}_{i,j}=\|\mathbf{r}_i-\mathbf{s}_j\|$  is only two (see analysis in Section \ref{sec:prosed-4}). This makes it impossible to solve for the four unknowns, leading to the widely accepted minimal configuration requirement: at least six/five/four sensors and four/five/six sources, respectively. However, if an alternative method can be found to bypass the principle related to the number of valid equations (known TOA measurements) and unknown locations, numerical solutions for the four unknowns should be feasible even with just four sensors and four sources for JSSL. Therefore, the boundaries of these four unknowns are derived in the next subsection.

\subsection{Triangle inequality for lower and upper boundaries of four unknowns}\label{sec:prosed-3}

In this subsection, before obtaining the solutions of four unknown variables $\hat{\alpha}$, $\hat{\beta}$, $\hat{\gamma}$ and $\hat{\eta}$, we derive both the lower and upper boundaries for those four unknowns by using the triangle inequality.

The triangle inequality states that in any triangle, the length of one side must be greater than or equal to the absolute difference between the lengths of the other two sides, and less than or equal to their sum. Specifically, for a triangle with side length $a_1$, $a_2$ and $a_3$, respectively, it holds that $|a_2-a_3| \leq a_1 \leq|a_2+a_3|$, where $|\bullet|$ denotes the absolute value. Therefore, from Figs. \ref{Con-fig2}(a) and (b), we can have the lower and upper boundaries for the four unknown variables $\hat{\alpha}$, $\hat{\beta}$, $\hat{\gamma}$ and $\hat{\eta}$ 
\begin{equation}
    \label{Con-16}
    \begin{cases}
 max(|\mathbf{d}_{1,j}-\mathbf{d}_{2,j}|) \leq \hat{\alpha} \leq min(|\mathbf{d}_{1,j}+\mathbf{d}_{2,j}|) \\
    max(|\mathbf{d}_{i,1}-\mathbf{d}_{i,2}|) \leq \hat{\beta} \leq min(|\mathbf{d}_{i,1}+\mathbf{d}_{i,2}|) \\   
     max(|\mathbf{d}_{i,1}-\mathbf{d}_{i,3}|) \leq \hat{\gamma} \leq min(|\mathbf{d}_{i,1}+\mathbf{d}_{i,3}|) \\
  max(|\mathbf{d}_{i,1}-\mathbf{d}_{i,4}|) \leq \hat{\eta} \leq min(|\mathbf{d}_{i,1}+\mathbf{d}_{i,4}|)            
    \end{cases},
\end{equation}
where $i$ ranges from $1$ to $M$ and $j$ ranges from $1$ to $N$. In the next subsection, we will show the alternative way to obtain the numerical solutions of four unknowns within the boundaries in Eq. (\ref{Con-16}) even the number of both sensors and sources is four.

\subsection{Numerical solutions for JSSL}
\label{sec:prosed-4}

\begin{table}[!t]
\renewcommand{\arraystretch}{0.45}
\begin{tabular}{l}
\hline
\textbf{\textit{Algorithm:} Numerical solutions for JSSL}
\\

\hline
\textbf{Input:} 
1.  $M$ and $N$;  \\
\qquad \quad 2. TOA measurements between sensors and sources. \\
\\ \textbf{Output:} Locations of $M$ sensors and $N$ sources. \\
\hline
\textbf{Step 1:} Obtain the lower and upper boundaries for $\hat{\alpha}$, $\hat{\beta}$, $\hat{\gamma}$ and $\hat{\eta}$  \\
\qquad \quad \quad with Eq.(\ref{Con-16}). \\

\textbf{Step 2:} Exclude the candidates of $\hat{\alpha}$, $\hat{\beta}$, $\hat{\gamma}$ and $\hat{\eta}$ in \textbf{Step 1}. \\
\textbf{Step 2.1:} Exclude the candidates of $\hat{\alpha}$ with $\mathbf{s}_{2,1}>0$ in Eq. (\ref{Con-4}). \\
\textbf{Step 2.2:} Exclude the candidates of $\hat{\beta}$, $\hat{\gamma}$ and $\hat{\eta}$ with Eq. (\ref{Con-14}): \\ \qquad \qquad $(\mathbf{C}_{1,:}\mathbf{V}_{:,1})^2\geq 0$,  $(\mathbf{C}_{1,:}\mathbf{V}_{:,2})^2 \geq 0$ and $(\mathbf{C}_{1,:}V_{:,3})^2 \geq 0$. \\

\textbf{Step 3:}  Test remaining candidates for $\hat{\alpha}$, $\hat{\beta}$, $\hat{\gamma}$ and $\hat{\eta}$: \\
\textbf{Step 3.1:} 
Obtain values for $\mathbf{s}_{2,1}$, $\mathbf{s}_{3,j}$, $\mathbf{C}_{3,:}$, $\mathbf{C}_{2,:}$ and $\mathbf{C}_{1,:}$  \\
\qquad \qquad with Eqs. (\ref{Con-3}), (\ref{Con-4}), (\ref{Con-7}), (\ref{Con-11}) (\ref{Con-15}); \\

\textbf{Step 3.2:} Obtain locations for all sensors and source \\
\qquad \qquad with Eqs. (\ref{Con-1}); \\
\textbf{Step 3.3:} Calculate the $Er$ in Eq. (\ref{Con-20}).\\
\textbf{Step 4:}  Choose the optimal value for $\hat{\alpha}$, $\hat{\beta}$, $\hat{\gamma}$ and $\hat{\eta}$ \\
\qquad \qquad with minimal value of $Er$ in \textbf{Step 3}. \\
\textbf{Step 5:} Obtain the locations for all sensor and sources \\
\qquad \qquad using \textbf{Step 3.1}  and \textbf{Step 3.2}.
\\
\hline
\end{tabular}
\end{table}

In this subsection, the numerical solutions for four unknowns $\hat{\alpha}$, $\hat{\beta}$, $\hat{\gamma}$ and $\hat{\eta}$ are derived by using the boundaries of those four unknowns in Eq. (\ref{Con-16}). Before estimating the numerical solutions for those four unknowns, we shall show the number of valid equations for clarifying the corresponding solutions of proposed numerical method. 

Since the distance $\mathbf{d}_{i,j}$ between $i^{th}$  sensor and $j^{th}$ source is formulated as $\|\mathbf{r}_i-\mathbf{s}_j\|$, we can divide $\mathbf{d}_{i,j}=\|\mathbf{r}_i-\mathbf{s}_j\|$ ($i=\begin{matrix}
    1, & \cdots, & M
\end{matrix}$ and $j=\begin{matrix}
    1, & \cdots, & N
\end{matrix}$) as three cases
\begin{small}
\begin{equation}
    \label{Con-17}
    \begin{cases}
           \mathbf{d}_{1,j}^2=\mathbf{r}_1^T\mathbf{r}_1+\mathbf{s}_j^T\mathbf{s}_j-2\mathbf{r}_1^T\mathbf{s}_j   \quad j= 
        1, \cdots, N \\
    \mathbf{d}_{i,1}^2=\mathbf{r}_i^T\mathbf{r}_i+\mathbf{s}_1^T\mathbf{s}_1-2\mathbf{r}_i^T\mathbf{s}_1   \quad i= 
        1, \cdots, M   \\
  \mathbf{d}_{i,j}^2=\mathbf{r}_i^T\mathbf{r}_i+\mathbf{s}_j^T\mathbf{s}_j-2\mathbf{r}_i^T\mathbf{s}_j   \quad i= 
        2, \cdots, M; j= 
        2, \cdots, N
    \end{cases}.
\end{equation}
\end{small}

Upon inspection of Eq. (\ref{Con-17}), since $\mathbf{r}_1=\begin{bmatrix}
    0, & 0, &0
\end{bmatrix}^T$, we can have $\mathbf{d}_{1,j}^2=\mathbf{s}_j^T\mathbf{s}_j$ for  $j= 1, \cdots, N$. Therefore, with $\mathbf{d}_{1,1}^2=\mathbf{s}_1^T\mathbf{s}_1$, we can have $\mathbf{d}_{i,1}^2=\mathbf{r}_i^T\mathbf{r}_i+\mathbf{d}_{1,1}^2-2\mathbf{r}_i^T\mathbf{s}_1$ for  $i= 
        1, \cdots, M$. Similarity, with $\mathbf{d}_{1,j}^2=\mathbf{s}_j^T\mathbf{s}_j$ and $-2\mathbf{r}_i^T(\mathbf{s}_j-\mathbf{s}_1)=\mathbf{U}_{i-1,:}\mathbf{V}_{:,j-1}=\mathbf{d}_{i,j}^2-\mathbf{d}_{i,1}^2-\mathbf{d}_{1,j}^2+\mathbf{d}_{1,1}^2$ (see Eqs. (\ref{pfeq7}), (\ref{preq3}) and (\ref{Con-1})), we can rewrite  $\mathbf{d}_{i,j}^2=\mathbf{r}_i^T\mathbf{r}_i+\mathbf{s}_j^T\mathbf{s}_j-2\mathbf{r}_i^T\mathbf{s}_j $ ($i=
        2,  \cdots, M$; $j= 
        2, \cdots, N$) in Eq. (\ref{Con-17}) as
    \begin{equation}
    \label{Con-18}
      \mathbf{d}_{i,1}^2-\mathbf{d}_{1,1}^2 = \mathbf{r}_i^T\mathbf{r}_i-2\mathbf{r}_i^T\mathbf{s}_1, 
    \end{equation}
which is same as the variant of $\mathbf{d}_{i,1}^2=\mathbf{r}_i^T\mathbf{r}_i+\mathbf{s}_1^T\mathbf{s}_1-2\mathbf{r}_i^T\mathbf{s}_1$ in Eq. (\ref{Con-17}) when $i=2, \cdots, M$.
Hence, we can reduce the redundancy in  Eq. (\ref{Con-17}) as
\begin{equation}
    \label{Con-19}
    \begin{cases}
          \mathbf{d}_{1,j}^2=\mathbf{s}_j^T\mathbf{s}_j
          \quad j= 
        1, \cdots, N \\
    \mathbf{d}_{i,1}^2-\mathbf{d}_{1,1}^2 = \mathbf{r}_i^T\mathbf{r}_i-2\mathbf{r}_i^T\mathbf{s}_1  \quad i= 
        1, \cdots, M 
    \end{cases}.
\end{equation}

Next, we shall show the valid equations in Eq. (\ref{Con-19}) for obtaining the numerical solutions of four unknowns. Since $\mathbf{r}_1=\begin{bmatrix}
    0, & 0, & 0
\end{bmatrix}^T$, the equation $ \mathbf{d}_{i,1}^2-\mathbf{d}_{1,1}^2 = \mathbf{r}_i^T\mathbf{r}_i-2\mathbf{r}_i^T\mathbf{s}_1$ is invalid when $i=1$. In addition, when $i=2$, $\mathbf{d}_{i,1}^2-\mathbf{d}_{1,1}^2 = \mathbf{r}_i^T\mathbf{r}_i-2\mathbf{r}_i^T\mathbf{s}_1$ is also invalid since it is same as the equation $\hat{\alpha}^2=\mathbf{d}_{1,j}^2+\mathbf{d}_{2,j}^2-2(\mathbf{r}_1-\mathbf{s}_j)^T(\mathbf{r}_2-\mathbf{s}_j)$ in Eq. (\ref{Con-2}) for $j=1$. Therefore, there are just $M-2$ valid equations for $\mathbf{d}_{i,1}^2-\mathbf{d}_{1,1}^2 = \mathbf{r}_i^T\mathbf{r}_i-2\mathbf{r}_i^T\mathbf{s}_1$. Next, we shall present the number of valid equations for  $ \mathbf{d}_{1,j}^2=\mathbf{s}_j^T\mathbf{s}_j$ in Eq. (\ref{Con-19}), where $j= 1, \cdots, N$. When $j=1$, $ \mathbf{d}_{1,1}^2=\mathbf{s}_1^T\mathbf{s}_1$ is invalid  since it is already used to formulate variables $\mathbf{s}_{2,1}$ and $\mathbf{s}_{3,1}$ (see Eqs. (\ref{Con-3}) and (\ref{Con-4})). When $j=2$, $3$ and $4$, $ \mathbf{d}_{1,j}^2=\mathbf{s}_j^T\mathbf{s}_j$ are invalid since they are also already used to formulate three variables $\mathbf{C}_{1,:}$ (see Eqs. (\ref{Con-12}), (\ref{Con-13}), (\ref{Con-14}) and (\ref{Con-15})), thus there are just $N-4$ valid equations for $ \mathbf{d}_{1,j}^2=\mathbf{s}_j^T\mathbf{s}_j$ when $j= 1, \cdots, N$. With those analysis, it is obvious that the number of valid equations for obtaining the solutions of four unknowns is only $M+N-6$ with TOA measurements.

With the number of valid equations $M+N-6$ in Eq. (\ref{Con-19}), it is obvious that it is impossible to obtain either closed form solutions or iterative solutions for four unknowns with any optimization methods once 1) $M=4$ and $N=4$; 2) $M=4$ and $N=5$ and  3) $M=5$ and $N=4$. Fortunately, thanks for the boundaries that we derived in Eq. (\ref{Con-16}), the acquisition of the numerical solutions for four unknowns becomes possible. In more details, first, we can divide the boundaries of four unknowns into several values of candidates with a given small step. Second, with those given candidates of four unknowns, we can obtain several sets of locations pertaining to the sensors and sources. Finally, the numerical solutions for JSSL can be obtained by choosing the location sets of sensors and sources that results in the minimal error with the valid equations in Eq. (\ref{Con-19}):

\begin{small}
\begin{equation}
    \label{Con-20}
    Er=\sqrt{\sum_{i=3}^{M}(\mathbf{d}_{i,1}^2-\mathbf{d}_{1,1}^2 - \mathbf{r}_i^T\mathbf{r}_i+2\mathbf{r}_i^T\mathbf{s}_1)^2}+\sqrt{\sum_{j=5}^{N}(\mathbf{d}_{1,j}^2-\mathbf{s}_j^T\mathbf{s}_j)^2}.
\end{equation}
\end{small}

It should be noted that when $N<5$, the second term in the right hand of Eq. (\ref{Con-20}) should be discarded. \textbf{\textit{Algorithm}} presents the pseudo code for localizing both sensors and sources with four unknowns $\hat{\alpha}$, $\hat{\beta}$, $\hat{\gamma}$ and $\hat{\eta}$.

\subsection{Extensions of proposed numerical method}\label{sec:prosed-5}

To valid the proposed numerical method that the locations of sensors and sources can be confirmed by four unknowns, and facilitate the applications of JSSL, we extend proposed numerical method to five different scenarios for JSSL: 

1) One co-located sensor and source and one known distance between a pair of sensors: In this scenario,  one additional sensor $\mathbf{\check{r}}$ is co-located with the $1^{st}$ source $s_1$, then we can obtain the solutions of three variables $\hat{\beta}$, $\hat{\gamma}$ and $\hat{\eta}$ since they are the known TOA measurements between three sources ($\mathbf{s}_1$, $\mathbf{s}_2$ and $\mathbf{s}_3$) and the additional sensor  $\mathbf{\check{r}}$. In addition, once the distance between $1^{st}$ sensor $\mathbf{r}_1$ and  $2^{nd}$ sensor $\mathbf{r}_2$ is known, the value of variable $\hat{\alpha}$ also can be obtained. Thus,  the locations of all sensors and sources can be obtained with closed-form solutions in terms of the proposed numerical method. 

2) One co-located sensor and sources: In this scenario, one additional sensor $\mathbf{\check{r}}$ is co-located with the $1^{st}$ source $s_1$~\cite{4}, so that we can obtain the solutions for $\hat{\beta}$, $\hat{\gamma}$ and  $\hat{\eta}$. Thus, only one variable $\hat{\alpha}$ is unknown and the task of JSSL is to estimate the optimal numerical solution for $\hat{\alpha}$ under the corresponding lower and upper boundaries in Eq. (\ref{Con-16}) with proposed \textbf{\textit{Algorithm}} in Subsection \ref{sec:prosed-4}.

3) Two known distances for one pair of sensors and one pair of sources: In this scenario, two distances between any two sensors and any two sources are given. Without losing the generality, we assume the two known distances to be $1^{st}$ and $2^{nd}$ sensors, and $1^{st}$ and $2^{nd}$ sources, respectively, thus the solutions for $\hat{\alpha}$ and $\hat{\beta}$ are known. Therefore, with the proposed \textbf{\textit{Algorithm}} in Subsection \ref{sec:prosed-4}, the task of JSSL is to choose the optimal  numerical solutions for $\hat{\gamma}$ and $\hat{\eta}$ under the corresponding lower and upper boundaries in Eq. (\ref{Con-16}).

4) One known distance between any two sensors: In this scenario, one distance between any two sensors is given. Without losing the generality, we assume the distance between $1^{st}$ and $2^{nd}$ sensors to be known, thus the solutions for $\hat{\alpha}$ can be obtained. Then the task of JSSL is to choose the optimal solutions for $\hat{\beta}$, $\hat{\gamma}$ and $\hat{\eta}$ under the corresponding lower and upper boundaries in Eq. (\ref{Con-16}). With this setup, it is clear that the number of known TOA measurements between sensors and sources is $MN$ and the number of unknowns pertaining to the locations of sensors and sources is $3(M+N)-7$ as $\hat{\alpha}=\mathbf{r}_{3,2}$. Once we apply the principle to JSSL that the number of equations should be larger than or equal to the number of unknowns, it implies $(M-3)(N-3)\geq 2$, indicating that four sensors and four sources are insufficient for the task of JSSL. However, with our numerical method shown in the \textbf{\textit{Algorithm}} (see Subsection \ref{sec:prosed-4}), by estimating the numerical solutions for the three unknowns $\hat{\beta}$, $\hat{\gamma}$ and $\hat{\eta}$,  
 we shall show the possibilities that four sensors and four source are sufficient for JSSL, relaxing the minimal configuration of state-of-the-arts during the past decades, facilitating the task of JSSL when the number of sensors and sources is not sufficient for localizing both sensors and sources by using traditional methods in state-of-the-arts.

 5) No prior information: In this scenario, there are not any known distance measurements  for any pairs of sensors and any pairs of sources. Thus the task of JSSL is to estimate the optimal solutions for four unknowns $\hat{\alpha}$, $\hat{\beta}$, $\hat{\gamma}$ and $\hat{\eta}$ under the corresponding lower and upper boundaries in Eq. (\ref{Con-16}). With this setup, it is obvious that the number of known TOA measurements between sensors and sources is $MN$, and the number of unknowns pertaining to the locations of sensors and sources is $3(M+N)-6$, resulting in the corresponding conditions for JSSL, i.e., $(M-3)(N-3)\geq 3$, implying the following configurations are insufficient for localizing both sensors and sources: 1) $M=4$ and $N=4$;   2) $M=4$ and $N=5$; 3) $M=5$ and $N=4$. However,  with our numerical method stated in the \textbf{\textit{Algorithm}} (see Subsection \ref{sec:prosed-4}),  the solutions of four unknowns can be estimated under the three minimal configurations above,  relaxing minimal configurations of state-of-the-arts in the past decades by reducing the number of sensors and sources for JSSL.   

 \section{Synthetic Experiments and Evaluations}\label{cha:over-exp}
The experimental settings are discussed in Subsection \ref{cha:over-exp-1}. Then we show the results of proposed numerical method in Subsections   \ref{cha:over-exp-2} and \ref{cha:over-exp-3}. In addition, by adding the Gaussian noise to both range measurements $\mathbf{d}_{i,j}$ and four unknowns with zero mean and a standard deviation $\sigma=\{10^{-6}, 10^{-4}, 10^{-3}, 10^{-2}\}$ $m$~\cite{sota-20}, the robustness of proposed numerical method can be shown. Finally, in Subsection \ref{cha:over-exp-4}, the limitation of proposed numerical method is illustrated. 

\subsection{Setup}
\label{cha:over-exp-1}

\subsubsection{Simulation data} The location of sensors and sources have been randomly generated by MATLAB under the uniform distribution within different room size of $1 \times 1 \times 1$ $m^3$~\cite{sota-20}, respectively. Without loosing the generality, the speed of sound is set to be $340 \ m/s$~\cite{sota-20}. Thus, the TOA measurements between sensors and sources can be generated by the locations of sensors and sources.
\subsubsection{Evaluation metric}
To evaluate and validate proposed numerical method, we compare the localization errors pertaining to the ground truth and estimation value of locations of sensors and sources, i.e.,
\begin{equation}
    \label{Con-21}
    EM=\frac{\sum_{i=1}^{M}\|\mathbf{r}_i-\mathbf{\hat{r}}_i\|+\sum_{j=1}^{N}\|\mathbf{s}_j-\mathbf{\hat{s}}_j\|}{M+N},
\end{equation}
where $\mathbf{\hat{r}}_i$ and $\mathbf{\hat{s}}_j$ are estimation location for $i^{th}$ sensor and $j^{th}$ source, respectively.

\subsection{Results for different configurations and noise intensity $\sigma$}\label{cha:over-exp-2}

 \begin{figure*}[!t]
    \includegraphics[trim=1.2cm 5.5cm 0.2cm 5.5cm, clip=true, width=0.95\textwidth ]{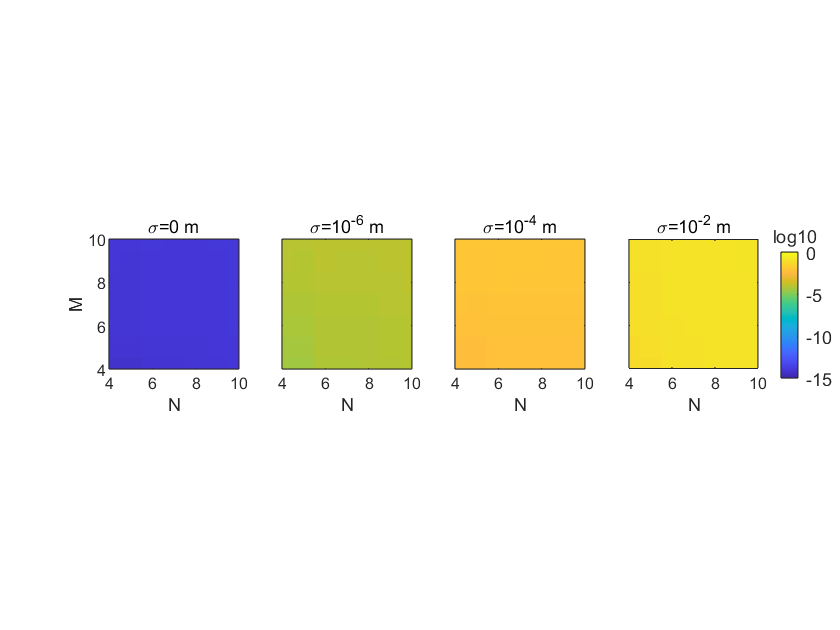}
    \caption{Average localization error $EM$ in Eq. (\ref{Con-21}) for proposed numerical method: scenario with one co-located additional sensor and source and one known distance between a pair of sensors ($\log10$ denotes the transformation of  $\log_{10}$; known $\hat{\alpha}$, $\hat{\beta}$, $\hat{\gamma}$ and $\hat{\eta}$).}
        \label{Con-fig3}
\end{figure*}

\begin{figure*}[!t]
\centering
\subfigure[One co-located additional sensor and source ($M=4$ and $N=4$): the principle for searching $\hat{\alpha}$ by using $Er$ in Eq. (\ref{Con-20}) (left), and the effect of step size for searcing $\hat{\alpha}$ on localization errors $EM$ in Eq. (\ref{Con-21}) (right).]{  \includegraphics[trim=0.5cm 3cm 0.5cm 4cm, clip=true, width=0.8\textwidth ]{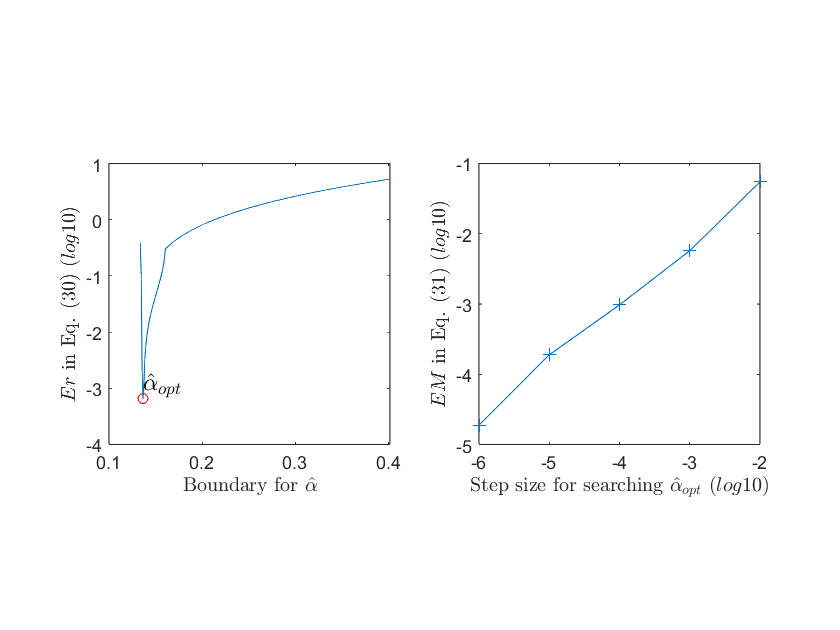}}

\subfigure[Average localization error $EM$ in Eq. (\ref{Con-21}) for proposed numerical method: scenario with one co-located additional sensor and source.]{ \includegraphics[trim=1.2cm 5.5cm 0.2cm 5.5cm, clip=true, width=0.95\textwidth ]{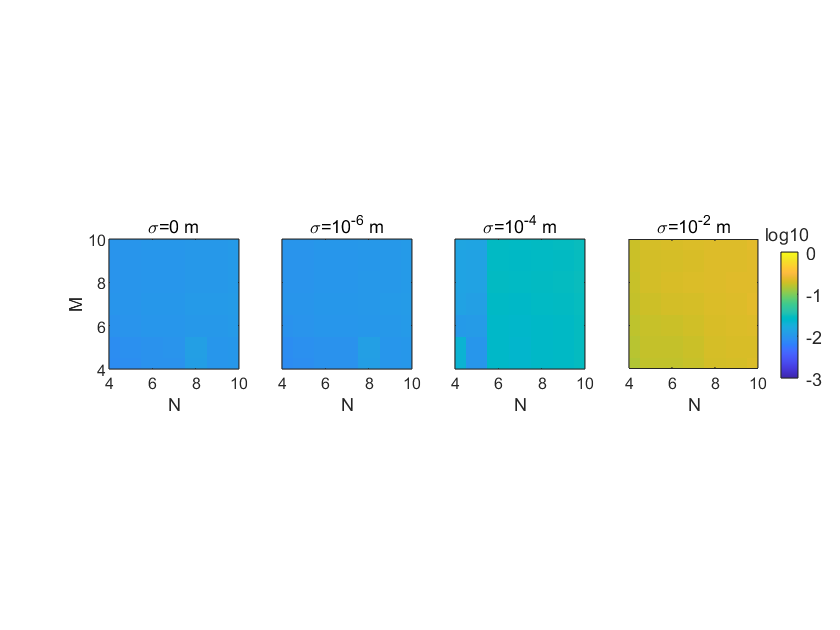}}

\subfigure[Average localization error $EM$ in Eq. (\ref{Con-21}) for proposed numerical method: scenario with two known distances of one pair of sensors and one pair of sources.]{ \includegraphics[trim=1.2cm 5.5cm 0.2cm 5.5cm, clip=true, width=0.95\textwidth ]{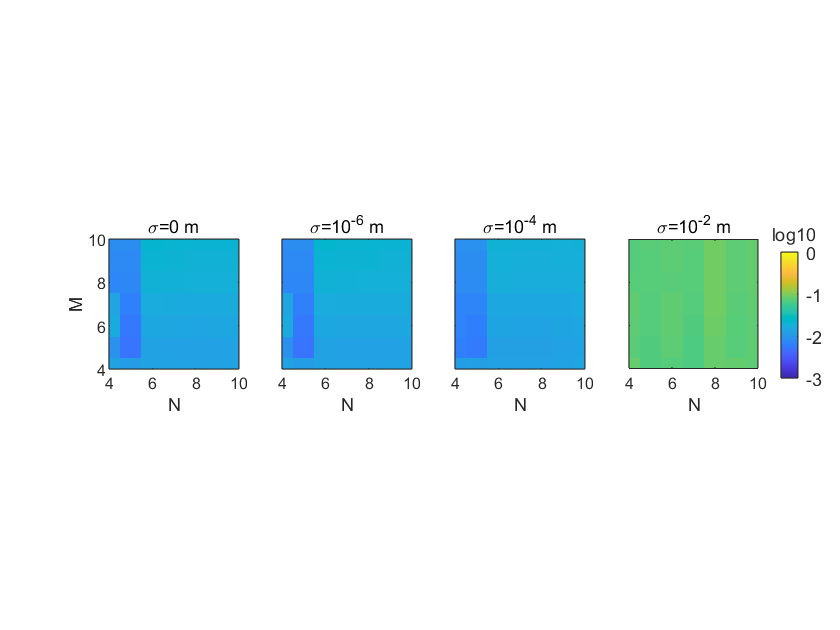}}
\caption{Results for proposed numerical method; $\log10$ denotes the transformation of $\log_{10}$ for the corresponding values; known $\hat{\beta}$, $\hat{\gamma}$ and $\hat{\eta}$ for figure (a); known $\hat{\beta}$, $\hat{\gamma}$ and $\hat{\eta}$ for figure (b); known $\hat{\alpha}$ and $\hat{\beta}$ for figure (c).}
\label{Con-fig4}
\end{figure*}

\begin{figure*}[!h]

\subfigure[Average localization error $EM$ in Eq. (\ref{Con-21}) for proposed numerical method: scenario with one known distance for a pair of sensors ($M=4$ and $N=4$ challenges the consensus of minimal configuration of state-of-the-arts).]{ \includegraphics[trim=1.2cm 5.5cm 0.2cm 5.5cm, clip=true, width=0.95\textwidth ]{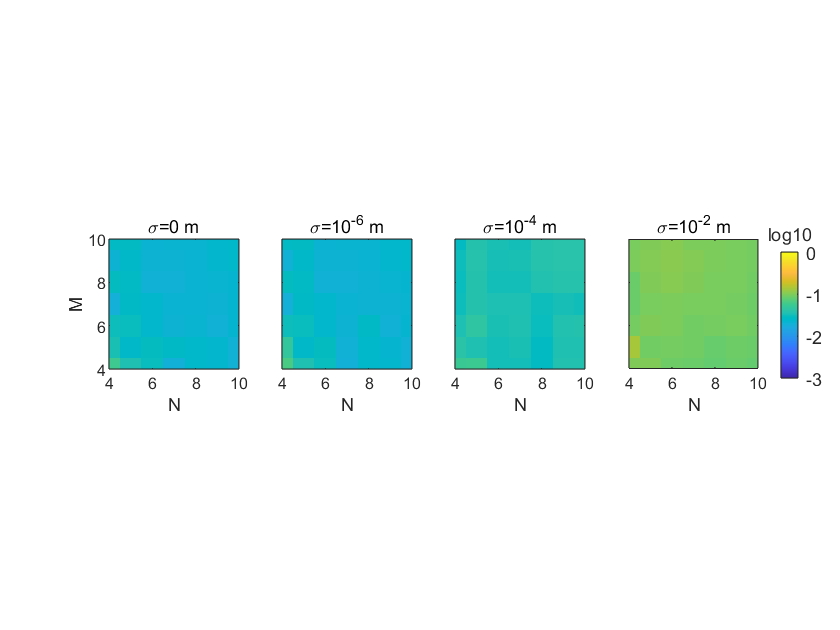}}

\subfigure[Average localization error $EM$ in Eq. (\ref{Con-21}) for proposed numerical method: scenario without known distances for any pairs of sensors and any pairs of sources ($M=4$ and $N=4$, $M=4$ and $N=5$ and $M=5$ and $N=4$ challenge the consensus of minimal configurations of state-of-the-arts).]{ \includegraphics[trim=1.2cm 5.5cm 0.2cm 5.5cm, clip=true, width=0.95\textwidth ]{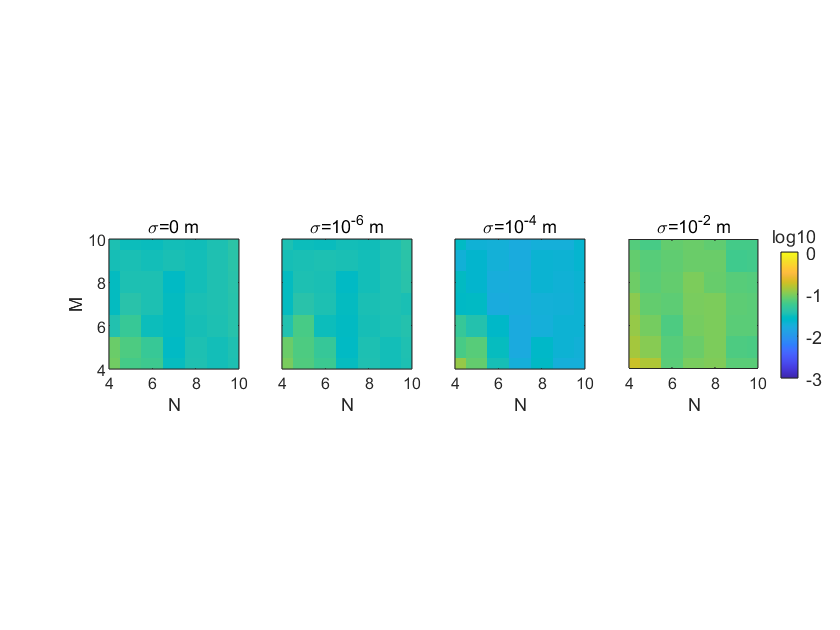}}
\caption{Results for two scenarios: unknown three variables $\hat{\beta}$, $\hat{\gamma}$ and $\hat{\eta}$ for figure (a); unknown four variables $\hat{\alpha}$, $\hat{\beta}$, $\hat{\gamma}$ and $\hat{\eta}$ for figure (b) ($\log10$ denotes the transformation of $\log_{10}$ for the corresponding values).}
\label{Con-fig5}
\end{figure*}

In this part, we run 30 different random configurations, given $M$ sensors and $N$ sources, then we average the localization errors $EM$ in Eq. (\ref{Con-21}) for 30 different configurations. Figs. \ref{Con-fig3}, \ref{Con-fig4} and \ref{Con-fig5} show the results with simulation data of 30 different configurations. Specifically, Figs. \ref{Con-fig3} and \ref{Con-fig4} display the results for first three scenarios in Section \ref{sec:prosed-5}, validating the statement of proposed numerical method that the locations of sensors and sources can be represented by four unknown variables once $M\geq 4$ and $N\geq 4$. While in Fig. \ref{Con-fig5}, we show the results for the last two scenarios in  Section \ref{sec:prosed-5}, not only validating the statement of our method that four variables can represent the locations of all sensors and sources, but also relaxing the minimal configurations in state-of-the-arts during the past decades, making the configurations of JSSL more flexible.

First, Fig. \ref{Con-fig3} shows the results for scenario with one co-located additional sensor and source and one known distance between a pair of sensors, it implies that four variables $\hat{\alpha}$, $\hat{\beta}$, $\hat{\gamma}$ and $\hat{\eta}$ in our method are known. As can be seen from  Fig. \ref{Con-fig3}, if the noise intensity is $0$ $m$, the average localization error $EM$ of 30 different configurations is about $10^{-15}$ $m$ when both $M$ and $N$ vary from $4$ to $10$, this validates the statement of proposed numerical method that the locations of all sensors and sources can be represented by four unknown variables. In addition, once the noise is introduced  into both four variables ($\hat{\alpha}$, $\hat{\beta}$, $\hat{\gamma}$ and $\hat{\eta}$) and the ranges $\mathbf{d}_{i,j}$ between sensors and sources, the average localization errors of 30 different configurations vary with the noise intensity $\sigma$. Specifically, if $\sigma=10^{-6}$ $m$, the average localization error $EM$ for 30 different configurations ranges from $10^{-5}$ $m$ to $10^{-4}$ $m$ when both $M$ and $N$ vary from $4$ to $10$. If  $\sigma=10^{-4}$ $m$, the average localization error $EM$ for 30 different configurations ranges from $10^{-3}$ $m$ to $10^{-2}$ $m$ when both $M$ and $N$ vary from $4$ to $10$. In addition, if  $\sigma=10^{-2}$ $m$, the average localization error $EM$ for 30 configurations ranges from $10^{-2}$ $m$ to $10^{-1}$ $m$ when both $M$ and $N$ vary from $4$ to $10$. In general, the results in Fig. \ref{Con-fig3} validate the statement of proposed numerical method that the locations of all sensor and sources can be represented by four unknown variables once $M\geq 4$ and $N\geq 4$, resulting in the statement that proposed numerical method can be applied to the scenario where one additional sensor and one source is co-located and the distance between a pair of sensors is known.

Second, Fig. \ref{Con-fig4} shows the results for other two scenarios, i.e., one co-located additional sensor and source, and two known distances between any two sensors and any two sources.
To setup the step size for searching the corresponding four unknowns $\hat{\alpha}$, $\hat{\beta}$, $\hat{\gamma}$ and $\hat{\eta}$, we visualize the results for the scenario of one co-located additional sensor and source with $M=4$ and $N=4$, as can be seen from the sub-figure of Fig. \ref{Con-fig4}(a) (left side), once we set the step size of $10^{-3}$ $m$ for searching the optimal $\hat{\alpha}$ within the boundaries in Eq. (\ref{Con-16}), the optimal $\hat{\alpha}$ can be obtained by choosing the minimal $Er$ in Eq. (\ref{Con-20}). Then from the another sub-figure of Fig. \ref{Con-fig4}(a) (right side), when the searching step of $\hat{\alpha}$ varies from $10^{-6}$ $m$ to $10^{-2}$ $m$, the value of average localization error $EM$ for 30 different configurations is nearly linear. Specifically, if the searching step of $\hat{\alpha}$ is $10^{-3}$ $m$, the average localization errors $EM$ for 30 configurations is only about  $10^{-2}$ $m$. Therefore, by considering the balance between  localization accuracy and time consuming for searching unknowns, if there is not specific remarks, we set the step size for searching four unknowns, $\hat{\alpha}$, $\hat{\beta}$, $\hat{\gamma}$ and $\hat{\eta}$, as $10^{-3}$ $m$ within the boundaries of Eq. (\ref{Con-16}). 

With the length of searching step of $10^{-3}$ $m$ for $\hat{\alpha}$, Fig. \ref{Con-fig4}(b)  shows the results for the scenario with one co-located additional sensor and source, it implies that three variables $\hat{\beta}$, $\hat{\gamma}$ and $\hat{\eta}$ are known. As can be seen from  Fig. \ref{Con-fig4}(b), if the noise intensity is $0$ $m$, the average localization error $EM$ of 30 different configurations varies from $0.008$ $m$ to $0.012$ $m$ when both $M$ and $N$ vary from $4$ to $10$, this also validates the statement of proposed numerical method that the locations of all sensors and sources can be represented by four unknown variables. In addition, once we add the noise to both $\hat{\beta}$, $\hat{\gamma}$, $\hat{\eta}$ and the ranges $\mathbf{d}_{i,j}$ between all sensors and sources, the average localization error varies with the noise intensity $\sigma$. Specifically, when $\sigma=10^{-6}$ $m$, the average localization error $EM$ for 30 configurations is similar with situation that  $\sigma=0$ $m$. If  $\sigma=10^{-4}$ $m$, the average localization error $EM$ for 30 configurations ranges from $0.009$ $m$ to $0.029$ $m$ when both $M$ and $N$ vary from $4$ to $10$.   If  $\sigma=10^{-2}$ $m$, the average localization error $EM$ for 30 configurations ranges from $0.14$ $m$ to $0.22$ $m$ when both $M$ and $N$ vary from $4$ to $10$. In general, the results in Fig. \ref{Con-fig4}(b) validate the statement of proposed numerical method that the locations of all sensor and sources can be represented by four unknown variables once $M\geq 4$ and $N\geq 4$, resulting in the conclusion that the proposed numerical method can be applied to the scenario where one additional sensor and one source is co-located, showing the step of $10^{-3}$ $m$ for searching $\hat{\alpha}$ is feasible for JSSL.  

With the length of searching step of $10^{-3}$ $m$ for both $\hat{\gamma}$ and $\hat{\eta}$, Fig. \ref{Con-fig4}(c)  shows the results for the scenario with two known distances between any one pair of sensors and one pair of sources, implying that two variables $\hat{\alpha}$ and $\hat{\beta}$ are known. As can be seen from  Fig. \ref{Con-fig4}(c), if the noise intensity is $0$ $m$, the average localization error $EM$ of 30 different configurations ranges from $0.005$ $m$ to $0.022$ $m$ when both $M$ and $N$ vary from $4$ to $10$, this also validates the statement of proposed numerical method that the locations of all sensors and sources can be represented by four unknown variables. In addition, once the noise is introduced to $\hat{\alpha}$, $\hat{\beta}$ and the ranges $\mathbf{d}_{i,j}$ between all sensors and sources, the average localization error $EM$ varies with the noise intensity $\sigma$. Specifically, if $\sigma=10^{-6}$ $m$ and $\sigma=10^{-4}$ $m$, the average localization errors $EM$ of those two situations for 30 configurations are similar with the results when  $\sigma=0$ $m$. In addition,   if  $\sigma=10^{-2}$ $m$, the average localization error $EM$ for 30 configurations ranges from $0.06$ $m$ to $0.08$ $m$ when both $M$ and $N$ vary from $4$ to $10$. In general, the results in Fig. \ref{Con-fig4}(c) validates the statement of proposed numerical method that the locations of all sensor and sources can be represented by four unknown variables once $M\geq 4$ and $N\geq 4$, resulting in the conclusion that proposed numerical method can be applied to the scenario when two distances between any one pair of sensors and any one pair of sources are known, showing the step of $10^{-3}$ $m$ for searching $\hat{\alpha}$ and $\hat{\beta}$ is feasible for JSSL. 

Besides, if $\sigma=10^{-2}$ $m$, by comparing the results from Figs. \ref{Con-fig4}(b) with  the results from Figs. \ref{Con-fig4}(c), it is obvious that the scenario of two known distances in Figs. \ref{Con-fig4}(c) obtains the better localization accuracy than the scenario of three known distances in Figs. \ref{Con-fig4}(b). In details, if the noises intensity in both ranges $\mathbf{d}_{i,j}$ and known variables is fixed ($\hat{\beta}$, $\hat{\gamma}$ and $\hat{\eta}$ for Fig. \ref{Con-fig4}(b) and $\hat{\gamma}$ and $\hat{\eta}$ for Fig. \ref{Con-fig4}(c)), by searching the optimal solutions for those remaining unknown variables ($\hat{\alpha}$ for Fig. \ref{Con-fig4}(b), $\hat{\alpha}$ and $\hat{\beta}$ for Fig. \ref{Con-fig4}(c)), the metric in Eq. (\ref{Con-20}) could choose the optimal values of unknown variables that leads the minimal $Er$ in Eq. (\ref{Con-20}), resulting in the better localization accuracy in Fig. \ref{Con-fig4}(c) than in Fig. \ref{Con-fig4}(b), as there are noises with three variables of $\hat{\beta}$, $\hat{\gamma}$ and $\hat{\eta}$ in Fig. \ref{Con-fig4}(b), but only two variables of $\hat{\gamma}$ and $\hat{\eta}$ in Fig. \ref{Con-fig4}(c) contain noises.

Third, Fig. \ref{Con-fig5}(a) and (b) shows the results for two scenarios: 1) one known distance of a pair of sensors; 2) no prior information for any pairs of sensors and any pairs of sources. With the length of searching step of $10^{-3}$ $m$ for three variables $\hat{\beta}$ $\hat{\gamma}$ and $\hat{\eta}$, Fig. \ref{Con-fig5}(a)  shows the localization results when $\hat{\alpha}$ is known. As can be seen from  Fig. \ref{Con-fig5}(a), if the noise intensity $\sigma$ is $0$ $m$, $10^{-6}$ $m$, $10^{-4}$ $m$ and $10^{-2}$ $m$, the average localization error $EM$ of 30 different configurations for different $\sigma$ is in the interval of  $\begin{bmatrix}
    0.01 & 0.05
\end{bmatrix}$ $m$, $\begin{bmatrix}
    0.01 & 0.06
\end{bmatrix}$ $m$, $\begin{bmatrix}
    0.02 & 0.06
\end{bmatrix}$ $m$ and $\begin{bmatrix}
    0.08 & 0.13
\end{bmatrix}$ $m$, respectively, this validates the statement of proposed numerical method that the locations of all sensors and sources can be represented by four  variables. More importantly, when $M=4$ and $N=4$, the average localization errors $EM$ of 30 different configurations for $\sigma=0$, $m$ $\sigma=10^{-6}$ $m$, $\sigma=10^{-4}$ $m$ and $\sigma=10^{-2}$ $m$ are $0.05$ $m$, $0.06$ $m$ $0.05$ $m$ and  $0.09$ $m$, respectively, this validates the statement  that proposed numerical method can be applied for localizing sensors and sources when $M=4$ and $N=4$, which is less than the consensus of minimal configurations of state-of-the-art over the past decades (see analysis of fourth scenario in Section \ref{sec:prosed-5}).

In addition, with the length of searching step of $10^{-3}$ $m$ for four unknown variables $\hat{\alpha}$, $\hat{\beta}$, $\hat{\gamma}$ and $\hat{\eta}$, Fig. \ref{Con-fig5}(b)  shows the corresponding localization results. As can be seen from  Fig. \ref{Con-fig5}(b), if the noise intensity $\sigma$ is $0$ $m$, $10^{-6}$ $m$, $10^{-4}$ $m$ and $10^{-2}$ $m$, the average localization errors $EM$ of 30 different configurations for different $\sigma$  are in the interval of  $\begin{bmatrix}
    0.02 & 0.09
\end{bmatrix}$ $m$, $\begin{bmatrix}
    0.02 & 0.09
\end{bmatrix}$ $m$, $\begin{bmatrix}
    0.01 & 0.11
\end{bmatrix}$ $m$ and $\begin{bmatrix}
    0.05 & 0.17
\end{bmatrix}$ $m$, respectively, this validates the proposed numerical method that the locations of all sensors and sources can be represented by four unknown variables. More importantly, when $M=4$ and $N=4$, the average localization errors $EM$ of 30 different configurations for $\sigma=0$ $m$, $\sigma=10^{-6}$ $m$, $\sigma=10^{-4}$ $m$ and $\sigma=10^{-2}$ $m$ are $0.09$ $m$, $0.09$ $m$, $0.11$ $m$ and  $0.17$ $m$, respectively. And when $M=4$ and $N=5$, the average localization error $EM$ of 30 different configurations for $\sigma=0$ $m$, $\sigma=10^{-6}$ $m$, $\sigma=10^{-4}$ $m$ and $\sigma=10^{-2}$ $m$ are $0.06$ $m$, $0.06$ $m$, $0.07$ $m$ and  $0.14$ $m$, respectively. Furthermore, when $M=5$ and $N=4$, the average localization error $EM$ of 30 different configurations for $\sigma=0$ $m$, $\sigma=10^{-6}$ $m$, $\sigma=10^{-4}$ $m$ and $\sigma=10^{-2}$ $m$ are $0.08$ $m$, $0.08$ $m$, $0.06$ $m$ and  $0.13$ $m$, respectively.
Those results from three cases above  validate that proposed numerical method can be applied for localizing both sensors and sources when the number of sensors and sources is less than the consensus of minimal configurations of state-of-the-art in the past decades (see analysis of fifth scenario in Section \ref{sec:prosed-5}).

\subsection{Visualization for localization under fixed configurations and noise intensity $\sigma$} \label{cha:over-exp-3}

\begin{figure*}[!t]
\centering
\subfigure[Known $\hat{\alpha}$; mean $EM$ for $\sigma=10^{-3}$ $m$: $0.04$ $m$.]{ \includegraphics[trim=0.2cm 0.5cm 0.2cm 0.5cm, clip=true, width=0.45\textwidth ]{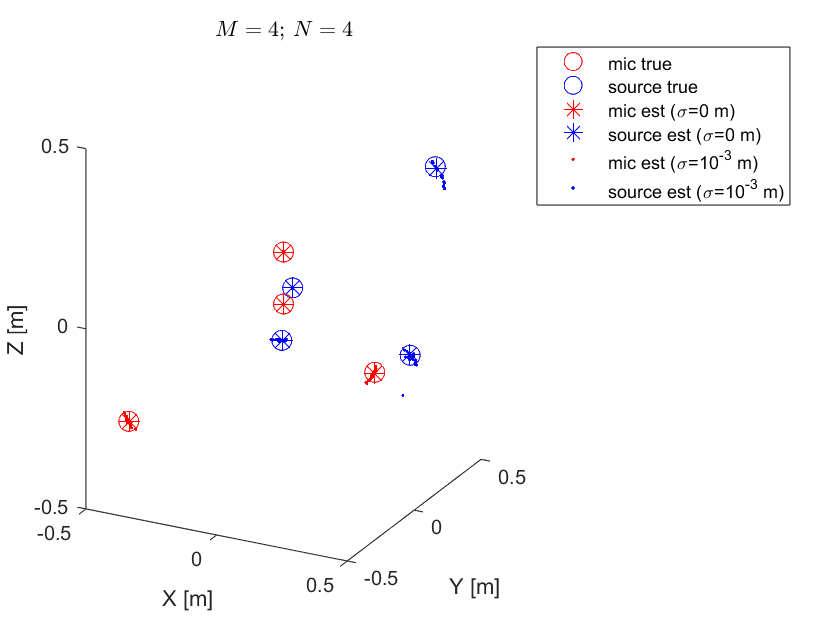}}
\subfigure[Unknown four variables; mean $EM$ for $\sigma=10^{-3}$ $m$: $0.10$ $m$.]{ \includegraphics[trim=0.2cm 0.5cm 0.2cm 0.5cm, clip=true, width=0.45\textwidth ]{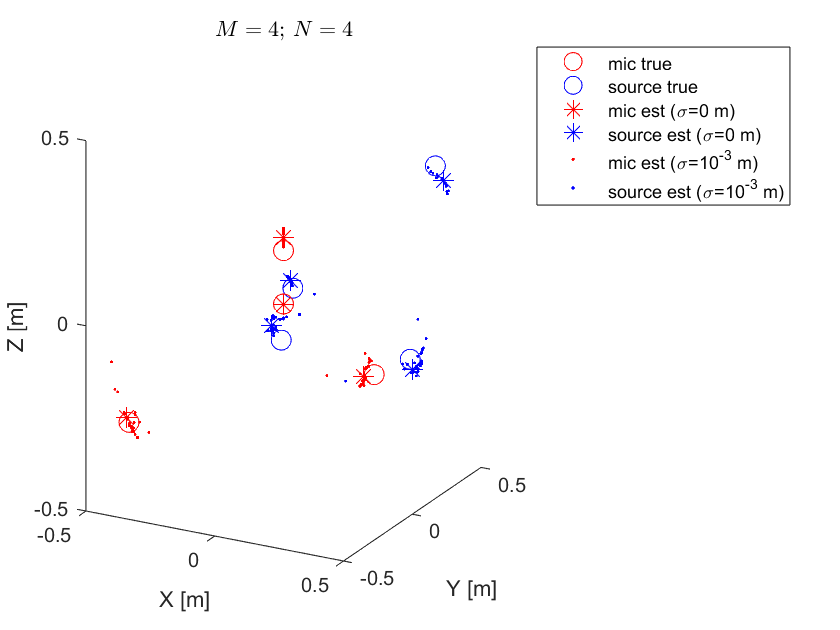}}

\subfigure[Unknown four variables; mean $EM$ for $\sigma=10^{-3}$ $m$: $0.06$ $m$.]{ \includegraphics[trim=0.2cm 0.5cm 0.2cm 0.5cm, clip=true, width=0.45\textwidth ]{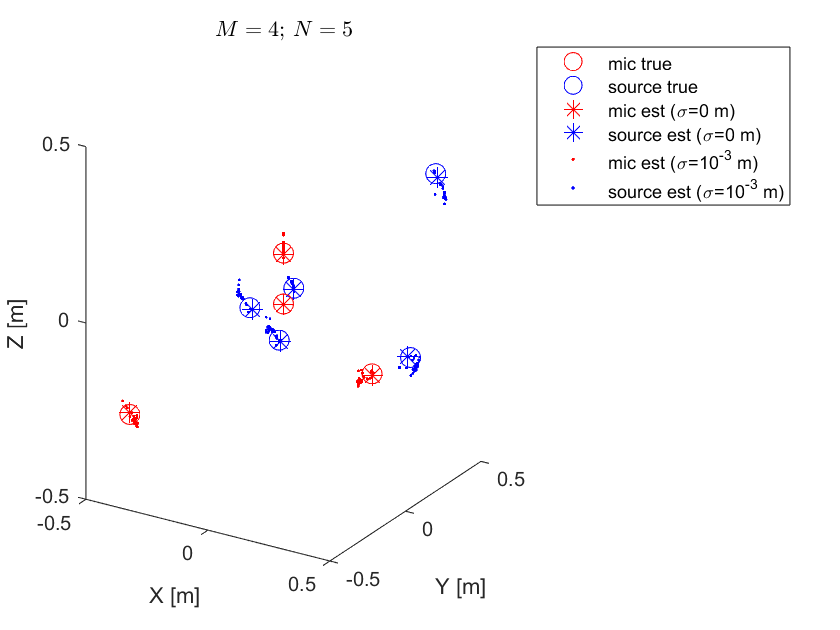}}
\subfigure[Unknown four variables; mean $EM$ for $\sigma=10^{-3}$ $m$: $0.04$ $m$]{ \includegraphics[trim=0.2cm 0.5cm 0.2cm 0.5cm, clip=true, width=0.45\textwidth ]{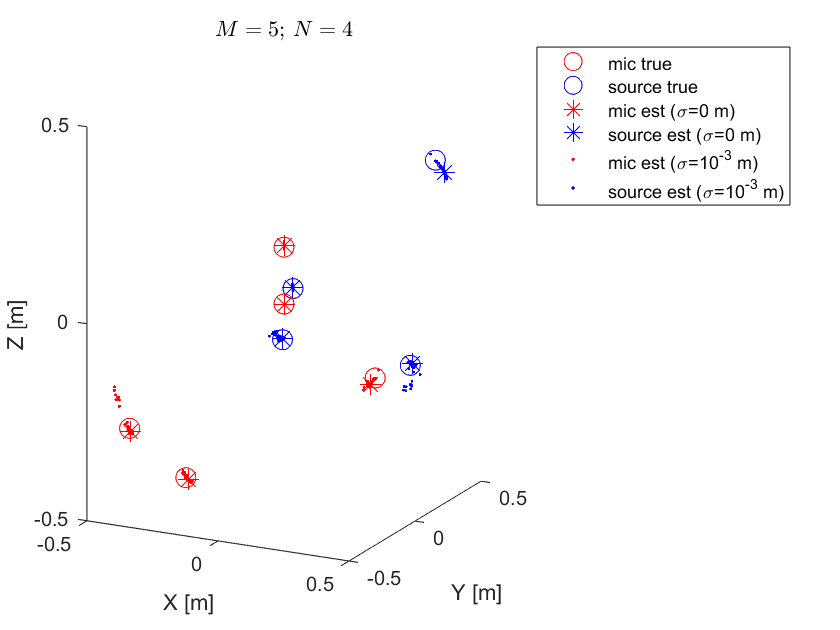}}
\caption{Visualization results of proposed numerical method for relaxing the minimal configurations of state-of-the-arts with the step size of $10^{-3}$ $m$.}
\label{Con-fig6}
\end{figure*}

In this part, to further validate and evaluate that the proposed numerical method can relax the minimal configurations for JSSL, we conduct 30 independent experiments by fixing the configurations and setting the noise intensity $\sigma=10^{-3}$, where the step size for searching unknowns is $10^{-3}$ $m$ within the boundaries in Eq. (\ref{Con-16}). Fig. \ref{Con-fig6} shows the results for the cases: 1) known $\hat{\alpha}$ with $M=4$ and $N=4$; 2) unknown four variables with $M=4$ and $N=4$; 3) unknown four variables with $M=4$ and $N=5$; 4) unknown four variables with $M=5$ and $N=4$. In addition, to visualize the impact of noise intensity of $\sigma=10^{-3}$ $m$ on localizing sensors and sources, we also display the ground truth and the locations of estimation with $\sigma=0$ $m$ for sensors and sources.

With the step size of $10^{-3}$ $m$ for searching three unknowns, Fig. \ref{Con-fig6}(a) shows the results for the case when $\hat{\alpha}$ is known. From Fig. \ref{Con-fig6}(a), we can see the $EM$ for $\sigma=0$ $m$ is $0.006$ $m$ where the localization accuracy is quite high. Once the noises are introduced into both $\hat{\alpha}$ and the distances between sensors and sources, the localization accuracy is still good, where the average localization error $EM$ for 30 different experiments is just $0.04$ $m$ when $\sigma=10^{-3}$ $m$. Then in Figs. \ref{Con-fig6}(b) (c) and (d), with the step size of $10^{-3}$ $m$ for searching four unknowns, we also show the results for the scenario of four unknowns with $M=4$ and $N=4$,  $M=4$ and $N=5$
 and $M=5$ and $N=4$, respectively. When $\sigma=0$ $m$,
 the localization errors $EM$ for $M=4$ and $N=4$,  $M=4$ and $N=5$ and $M=5$ and $N=4$ are $0.079$ $m$, $0.016$ $m$, $0.021$ $m$, respectively. Once the noises ($\sigma=10^{-3}$ $m$) are introduced into the distances $\mathbf{d}_{i,j}$ between sensors and sources, the localization accuracy is still good, where the average localization error $EM$ for 30 independent experiments is just $0.1$ $m$, $0.06$ $m$ and $0.04$ $m$ when $M=4$ and $N=4$,  $M=4$ and $N=5$ and $M=5$ and $N=4$, respectively. Therefore, those results from Fig. (\ref{Con-fig6}) validates the statement again that our method  relaxes the minimal configurations for JSSL, making the configurations of JSSL more flexible, contributing the realm of signal processing.



\subsection{Limitations}\label{cha:over-exp-4}

The proposed numerical method transforms the localization problems into the estimation of four unknowns and determines their lower and upper boundaries using triangle inequalities. The solutions for these unknowns are then obtained by utilizing the known TOA measurements between sensors and sources. This approach challenges the long-standing consensus of minimal configurations of state-of-the-art methods in the past decades, making  configurations of JSSL  more flexible.

In a simulation room of $1 \times 1 \times 1$ $m^3$
  with a search step of $10^{-3}$ $m$, the proposed numerical method achieved accurate localization results. Although these results are from simulations, the only requirement for the proposed method is having the TOA measurements between microphones and sources, suggesting it can also achieve accurate localization in real environments. Furthermore, the method is applicable to larger rooms, as room size is not a prerequisite. However, a noticeable limitation of the proposed numerical method is the time required to search the optimal solutions for unknowns. For example, if the boundaries for all four unknowns span one meter, the number of iterations required  for searching the optimal numerical solutions of JSSL is \(10^9\) and \(10^{12}\) when \(\hat{\alpha}\) is known and all four variables are unknown, respectively.

\section{Conclusion}
In this paper, the main focus is to relax the minimal configurations of state-of-the-art during past decades by using the TOA measurements between sensors and sources.
By formulating the localization problems with several triangles and applying the laws of cosine to those triangles, the localization problems of sensors and sources can be transformed to the estimation of four unknown distances pertaining to a pair of sensors and three pairs of sources. Then the triangle inequalities have been used for obtaining the lower and upper boundaries for the four unknown distances, so that we can search the optimal numerical solutions for those four unknown variables under the corresponding lower and upper boundaries, given an appropriate step size.  Experimental results validate the feasibility of the proposed numerical method: 1) the locations of all sensors and sources can be represented by four unknown variables when the number of both sensors and sources is greater than or equal to four; 2) four sensors and four sources are sufficient to localize both sensors and sources when the distance of one pair of sensors is known. 3) if there is no prior information pertaining to the distances of a pair of sensors and three pairs of sources, we can still localize all sensors and source when both the number of sensor and sources is four.  
Therefore, the research presented in this paper marks a significant milestone in the field of JSSL, as it challenges the longstanding assumption that JSSL requires a minimum of six/five/four sensors and four/five/six sources, respectively. Additionally, it introduces greater flexibility to JSSL configurations.

In addition, the research presented in this paper opens a new avenue for JSSL by challenging the longstanding consensus on minimal configurations. By breaking this consensus, we pave the way for exploring more efficient methods to solve for the four unknowns, establishing a new sub-topic within JSSL research. Furthermore, reducing the number of sensors and sources furtherly presents another promising direction for future studies in JSSL.

\bibliographystyle{IEEEtran}
\bibliography{Maindraft}

\begin{thebibliography}{10}
\providecommand{\url}[1]{#1}
\csname url@samestyle\endcsname
\providecommand{\newblock}{\relax}
\providecommand{\bibinfo}[2]{#2}
\providecommand{\BIBentrySTDinterwordspacing}{\spaceskip=0pt\relax}
\providecommand{\BIBentryALTinterwordstretchfactor}{4}
\providecommand{\BIBentryALTinterwordspacing}{\spaceskip=\fontdimen2\font plus
\BIBentryALTinterwordstretchfactor\fontdimen3\font minus
  \fontdimen4\font\relax}
\providecommand{\BIBforeignlanguage}[2]{{%
\expandafter\ifx\csname l@#1\endcsname\relax
\typeout{** WARNING: IEEEtran.bst: No hyphenation pattern has been}%
\typeout{** loaded for the language `#1'. Using the pattern for}%
\typeout{** the default language instead.}%
\else
\language=\csname l@#1\endcsname
\fi
#2}}
\providecommand{\BIBdecl}{\relax}
\BIBdecl

\bibitem{sota-20}
T.~K. Le and N.~Ono, ``Closed-form and near closed-form solutions for toa-based
  joint source and sensor localization,'' \emph{IEEE Transactions on Signal
  Processing}, vol.~64, no.~18, pp. 4751--4766, 2016.

\bibitem{Consensus-1}
T.~V. den Bogaert, S.~Doclo, J.~Wouters, and M.~Moonen, ``The effect of
  multimicrophone noise reduction systems on sound source localization by users
  of binaural hearing aids,'' \emph{The Journal of the Acoustical Society of
  America}, vol. 124, no. 484-497, p.~1, 2008.

\bibitem{Consensus-2}
Z.~Liu, ``Sound source seperation with distributed microphone arrays in the
  presence of clocks synchronization errors,'' \emph{In International Workshop
  for Acoustic Echo and Noise Control}, 2008.

\bibitem{gcc}
M.~S. Brandstein and H.~F. Silverman, ``A robust method for speech signal
  time-delay estimation in reverberant rooms,'' \emph{Proc. IEEE Int. Conf.
  Acoust. Speech, Signal Process.}, vol.~1, pp. 375--378, 1997.

\bibitem{phdrevis-1}
C.~Knapp and G.~Carter, ``The generalized correlation method for estimation of
  time delay,'' \emph{IEEE transactions on acoustics, speech, and signal
  processing}, vol.~24, no.~4, pp. 320--327, 1976.

\bibitem{phdrevis-2}
M.~Azaria and D.~Hertz, ``Time delay estimation by generalized cross
  correlation methods,'' \emph{IEEE Transactions on Acoustics, Speech, and
  Signal Processing}, vol.~32, no.~2, pp. 280--285, 1984.

\bibitem{phdrevis-3}
G.~A. Darbellay and I.~Vajda, ``Estimation of the information by an adaptive
  partitioning of the observation space,'' \emph{IEEE Transactions on
  Information Theory}, vol.~45, no.~4, pp. 1315--1321, 1999.

\bibitem{phdrevis-4}
A.~Kraskov, H.~Stögbauer, and P.~Grassberger, ``Estimating mutual
  information,'' \emph{Physical review E}, vol.~69, no.~6, p. 066138, 2004.

\bibitem{myself-1}
F.~Cao, Y.~Cheng, A.~M. Khan, Z.~Yang, S.~M.~A. Kazmi, and Y.~Chang, ``Low rank
  properties for synchronizing microphones and sources in ad-hoc wireless
  acoustic sensor network,'' \emph{IEEE Trans. Mob. Comput., Under Review},
  2024.

\bibitem{myself-2}
F.~Cao, Y.~Cheng, A.~M. Khan, Z.~Yang, and Z.~Guo, ``Are microphone signals
  alone sufficient for toa-based self-localization?'' \emph{Accepted at 29th
  International Conference on Automation and Computing (ICAC)}, 2024.

\bibitem{31}
T.~K. Le and N.~Ono, ``Closed-form and near closed-form solutions for
  tdoa-based joint source and microphone localization,'' \emph{IEEE Trans.
  Signal Process.}, vol.~65, no.~5, pp. 1207--1221, 2016.

\bibitem{phdrevis-5}
N.~D. Gaubitch, W.~B. Kleijn, and R.~Heusdens, ``Auto-localization in ad-hoc
  microphone arrays,'' \emph{In 2013 IEEE international conference on
  acoustics, speech and signal processing}, pp. 106--110, 2013.

\bibitem{phdrevis-6}
S.~Thrun, ``Affine structure from sound,'' \emph{Advances in Neural Information
  Processing Systems}, vol.~18, 2005.

\bibitem{phdrevis-7}
S.~H.~C. and A.~Z. Robinson, ``Passive source localization employing
  intersecting spherical surfaces from time-of-arrival differences,''
  \emph{IEEE Transactions on Acoustics, Speech, and Signal Processing},
  vol.~35, no.~8, pp. 1223--1225, 1987.

\bibitem{phdrevis-8}
J.~Smith and J.~Abel, ``Passive source localization employing intersecting
  spherical surfaces from time-of-arrival differences,'' \emph{IEEE
  Transactions on Acoustics, Speech, and Signal Processing}, vol.~35, no.~12,
  pp. 1661--1669, 1987.

\bibitem{phdrevis-9}
Y.~T. Chan and K.~C. Ho, ``A simple and efficient estimator for hyperbolic
  location,'' \emph{IEEE Transactions on Acoustics, Speech, and Signal
  Processing}, vol.~42, no.~8, pp. 1905--1915, 1994.

\bibitem{phdrevis-10}
L.~Yang and K.~C. Ho, ``An approximately efficient tdoa localization algorithm
  in closed-form for locating multiple disjoint sources with erroneous sensor
  positions,'' \emph{IEEE Transactions on Signal Processing}, vol.~57, no.~12,
  pp. 4598--4615, 2009.

\bibitem{phdrevis-11}
M.~S. Brandstein, J.~E. Adcock, and H.~F. Silverman, ``A closed-form location
  estimator for use with room environment microphone arrays,'' \emph{IEEE
  transactions on Speech and Audio Processing}, vol.~5, no.~1, pp. 45--50,
  1997.

\bibitem{5}
V.~C. Raykar, I.~V. Kozintsev, and R.~Lienhart, ``Position calibration of
  microphones and loudsources in distributed computing platforms,'' \emph{IEEE
  Trans. Speech, Audio Process.}, vol.~13, no.~1, pp. 70--83, 2004.

\bibitem{sota-3}
W.~S. Torgerson, ``Multidimensional scaling: I. theory and method,''
  \emph{Psychometrika}, vol.~17, no.~4, pp. 401--419, 1952.

\bibitem{phdrevis-14}
J.~R. I.~Dokmanic, R.~Parhizkar and M.~Vetterli, ``Euclidean distance matrices:
  essential theory, algorithms, and applications,'' \emph{IEEE Signal
  Processing Magazine}, vol.~32, no.~6, pp. 12--30, 2015.

\bibitem{phdrevis-12}
L.~Liberti and C.~Lavor, ``Euclidean distance geometry,'' \emph{Berlin:
  Springer}, vol.~3, 2017.

\bibitem{phdrevis-13}
J.~C. Gower, ``Properties of euclidean and non-euclidean distance matrices,''
  \emph{Linear algebra and its applications}, vol.~67, pp. 81--97, 1985.

\bibitem{4}
M.~Crocco, A.~D. Bue, and V.~Murino, ``A bilinear approach to the position
  self-calibration of multiple microphones,'' \emph{IEEE Trans. Signal
  Process.}, vol.~60, no.~2, pp. 660--673, 2011.

\bibitem{9}
P.~H. Schönemann, ``On metric multidimensional unfolding,''
  \emph{Psychometrika}, vol.~35, no.~3, pp. 349--366, 1970.

\bibitem{phdrevis-15}
T.~Le and N.~Ono, ``Numerical formulae for toa-based microphone and source
  localization,'' \emph{In 2014 14th International Workshop on Acoustic Signal
  Enhancement (IWAENC)}, pp. 178--182, 2014.

\bibitem{phdrevis-16}
T.~K. Le and N.~Ono, ``Reference-distance estimation approach for tdoa-based
  source and sensor localization,'' \emph{In 2015 IEEE International Conference
  on Acoustics, Speech and Signal Processing (ICASSP)}, pp. 2549--2553, 2015.

\bibitem{mfaoa1}
T.~K. Le and K.~C. Ho, ``Uncovering source ranges from range differences
  observed by sensors at unknown positions: Fundamental theory,'' \emph{IEEE
  Trans. Signal Process.}, vol.~67, no.~10, pp. 2665--2678, 2019.

\bibitem{30}
------, ``Algebraic complete solution for joint source and microphone
  localization using time of flight measurements,'' \emph{IEEE Trans. Signal
  Process.}, vol.~68, pp. 1853--1869, 2020.

\bibitem{sota-15}
M.~Pollefeys and D.~Nister, ``Direct computation of sound and microphone
  locations from time-difference-of-arrival data,'' \emph{2008 IEEE
  International Conference on Acoustics, Speech and Signal Processing}, pp.
  2445--2448, 2008.

\bibitem{sota-14}
A.~Karbasi, S.~Oh, R.~Parhizkar, and M.~Vetterli, ``Ultrasound tomography
  calibration using structured matrix completion,'' \emph{The 20th
  International Congress on Acoustics}, 2010.

\bibitem{simu1}
D.~E. Badawy, V.~Larsson, M.~Pollefeys, and I.~Dokmanic, ``Localizing
  unsynchronized sensors with unknown sources,'' \emph{IEEE Trans. Signal
  Process.}, vol.~71, pp. 641--654, 2023.

\bibitem{sota-27}
N.~D. Gaubitch, W.~B. Kleijn, and R.~Heusdens, ``Calibration of distributed
  sound acquisition systems using toa measurements from a moving acoustic
  source,'' \emph{2014 IEEE International Conference on Acoustics, Speech and
  SignalProcessing (ICASSP)}, pp. 7455--7459, 2014.

\bibitem{sota-11}
Z.~Liu, Z.~Zhang, L.~W. He, and P.~Chou, ``Energy-based sound source
  localization and gain normalization for ad hoc microphone arrays,'' \emph{in
  Proc. IEEE Int. Conf. Acoust. Speech, Signal Process.}, p. 761–764, 2007.

\bibitem{sota-1}
I.~McCowan, M.~Lincoln, and I.~Himawan, ``Microphone array shape calibration in
  diffuse noise fields,'' \emph{IEEE Transactions on Audio, Speech, and
  Language Processing}, vol.~16, no.~3, pp. 666--670, 2008.

\bibitem{sota-24}
N.~Ono, H.~Kohno, N.~Ito, and S.~Sagayama, ``Blind alignment of asynchronously
  recorded signals for distributed microphone array,'' \emph{2009 IEEE Workshop
  on Applications of Signal Processing to Audio and Acoustics}, pp. 161--164,
  2009.

\bibitem{sota-29}
Y.~Kuang and K.~Åström, ``Stratified sensor network self-calibration from
  tdoa measurements,'' \emph{21st European Signal Processing Conference}, pp.
  1--5, 2013.

\bibitem{Consensus-3}
E.~Bolker and B.~Roth, ``When is a bipartite graph a rigid framework?''
  \emph{Pacific Journal of Mathematics}, vol.~90, no.~1, pp. 27--44, 1980.

\bibitem{Consensus-5}
W.~Adams and P.~Loustaunau, ``An introduction to gröbner bases,''
  \emph{American Mathematical Society}, vol.~3, 2022.

\bibitem{Consensus-6}
D.~Grayson and M.~Stillman, ``Macaulay2,'' \emph{http://www.math.uiuc.edu/
  Macaulay2/}.

\bibitem{sota-17}
Y.~Kuang, S.~Burgess, A.~Torstensson, and K.~Åström, ``A complete
  characterization and solution to the microphone position self-calibration
  problem,'' \emph{2013 IEEE International Conference on Acoustics, Speech and
  Signal Processing}, pp. 3875--3879, 2013.

\bibitem{sota-18}
S.~Burgess, Y.~Kuang, and K.~Åström, ``Toa sensor network calibration for
  receiver and transmitter spaces with difference in dimension,'' \emph{21st
  European Signal Processing Conference (EUSIPCO 2013)}, pp. 1--5, 2013.

\bibitem{Consensus-4}
S.~Zhayida, S.~Burgess, Y.~Kuang, and K.~Åström, ``Minimal solutions for dual
  microphone rig self-calibration,'' \emph{In 2014 22nd European Signal
  Processing Conference}, pp. 2260--2264, 2014.

\bibitem{phdrevis-17}
G.~W. Stewart, ``On the early history of the singular value decomposition,''
  \emph{SIAM review}, vol.~35, no.~4, pp. 2549--2553, 1993.

\bibitem{rigidbi}
G.~Kalai, E.~Nevo, and I.~Novik, ``Bipartite rigidity,'' \emph{Transactions of
  the American Mathematical Society}, vol. 368, no.~8, pp. 5515--5545, 2016.

\end{thebibliography}

\end{document}